\documentstyle[aps,epsfig,amssymb]{revtex}

\newcommand{\module}[1] { {\mid\! #1\! \mid} }
\newcommand{\esp}       { \;\;\;\; }
\newcommand{\vect}[1]   { \mbox{\boldmath{$#1$}} }
\newcommand{\vsmall}[1] { {\scriptstyle{#1}} }
\newcommand{\smvec}[1]  { \vect{\vsmall{#1}} }
\newcommand{\refp}[1]   {(\ref{#1})}

\begin{document}
\draft
\title{A new construction for scalar wave equations in inhomogeneous 
media}
\author{S. {\sc de Toro Arias}\thanks{E-mail~: {\tt sdetoro@spec.saclay.cea.fr}}
and C. {\sc Vanneste}\thanks{Author for correspondence. 
E-mail~: {\tt vanneste@ondine.unice.fr}}}
\address{Laboratoire de Physique de la Mati\`{e}re Condens\'{e}e, 
CNRS UMR $6622$,\\
Universit\'{e} de Nice-Sophia Antipolis, \\
Parc Valrose, B. P. $71$, $06108$ Nice Cedex $2$, France}
\maketitle

\vspace{5mm}
\begin{flushleft}
J. Phys. I France {\bf 7} (1997) 1071--1096 \\
Shortened version of the title~:
{\bf scalar wave equations in inhomogeneous media}
\end{flushleft}

\vspace{5mm}
\begin{flushleft}
PACS~: $03.40$.Kf -- Waves and propagation~: general mathematical aspects \\
PACS~: $42.25$.Fx -- Diffraction and scattering \\
PACS~: $02.70$.-c -- Computational techniques
\end{flushleft}


\begin{abstract}
{\bf Abstract~:} The paper describes a formulation of 
discrete scalar wave propagation 
in an inhomogeneous medium by the use of elementary processes 
obeying a discrete Huygens' principle and satisfying fundamental 
symmetries such as time-reversal, reciprocity and isotropy. Its 
novelty is the systematic derivation of a unified equation which, 
properly tuned by a single parameter, leads to either the Klein-Gordon 
equation or the Schr\"{o}dinger equation. The generality of this 
method enables one to consider its extension to other types of 
discrete wave equations on any kind of discrete lattice. \\

{\bf R\'esum\'e~:} Cet article formule un mod\`{e}le discret de 
propagation d'ondes scalaires dans un milieu h\'{e}t\'{e}rog\`{e}ne \`{a} 
l'aide de processus  \'{e}l\'{e}mentaires qui ob\'{e}issent \`{a} un 
principe de Huygens discret et satisfont \`{a} certaines sym\'{e}tries 
fondamentales telles que le renversement temporel, la r\'{e}ciprocit\'{e} 
et l'isotropie. Son originalit\'{e} consiste en la d\'{e}rivation 
syst\'{e}matique d'une \'{e}quation unifi\'{e}e qui selon la 
valeur d'un seul param\`{e}tre conduit soit \`{a} l'\'{e}quation 
de Klein-Gordon, soit \`{a} l'\'{e}quation de Schr\"{o}dinger. 
La m\'{e}thode est suffisamment g\'{e}n\'{e}rale pour pouvoir 
envisager son extension \`{a} d'autres types d'\'{e}quations 
d'onde sur toute sorte de r\'{e}seau discret.
\end{abstract}


\section{Introduction}
The formulation of wave propagation by the use of Huygens' 
principle has stimulated extensive work in the past. Such an 
approach was first pioneered by P. B. Johns and coworkers who 
introduced the Transmission Line Matrix Modeling method 
(TLM)\cite{johns71,johns74a} to solve the Maxwell 
equations in large electromagnetic 
structures\cite{hoefer} which was later extended to the description of 
diffusive processes\cite{johns77}. Linear wave propagation is mediated 
by short voltage impulses which propagate along the bonds and 
are scattered on each node of a Cartesian mesh of transmission 
lines. The main idea behind such a formulation is a discrete 
equivalent of Huygens' principle, namely a principle of action-by-proximity 
obeyed by the pulses when they propagate at each time step from 
one node to the neighbor nodes and the emission of secondary 
wavelets at each node as described by the scattering process 
which radiates the incident energy in all directions. Besides 
the appealing viewpoint of Huygens' principle, the advantages 
of the method as a powerful tool for the numerical computation 
of wave propagation in complex and large structures have already 
been emphasized in the TLM method\cite{hoefer}. A similar approach in 
which the pulses were just scalar quantities propagating along 
the bonds of a Cartesian lattice, was recently advocated for 
studying time-dependent wave propagation in large inhomogeneous 
media\cite{vanneste,enders92,sebbah}.

Here, we focus on the nature of the wave equations which result 
from such a formulation, especially on the possibility of retrieving 
standard wave equations. This question has already been raised 
in previous works where the scalar wave equation, the Klein-Gordon 
equation and Maxwell equations have been exhibited. In the TLM 
method, the equivalence between the voltage and current impulses 
with the electric and magnetic fields of Maxwell equations was 
initially demonstrated on a two-dimensional mesh\cite{johns71}. The method 
was then extended to three dimensions and different features 
have been incorporated by many authors in order to describe losses, 
inhomogeneous media and boundary conditions\cite{hoefer}. Another recent 
formulation\cite{sornette} lead their authors to the classical wave equation, 
the Klein-Gordon equation and to the Schr\"{o}dinger equation. 
However, due to restricting assumptions, these last results were 
limited to scalar waves in homogeneous media, and the Schr\"{o}dinger 
equation was ill-defined at the continuum limit. 

This paper describes what is essentially a formulation from ``first 
principles'' of such an approach. The aim is to generalize the 
previous results to various kinds of waves in inhomogeneous media 
by considering the most basic properties obeyed by the scattering 
nodes. Such a formulation not only leads to the wave equations 
already mentioned but also results in a time-dependent Schr\"{o}dinger 
equation which is well defined.  Another advantage is the possible 
generalization of the method from the scalar waves considered 
in this paper to spinor\cite{sdetoro} or vector waves.

At each time step, the time-dependent discretized wave is defined 
as a linear combination of incident currents at each node of 
the Cartesian lattice. The nature of the wave depends on the 
number of distinct propagating currents on each bond of a given 
node. For example, one or several kinds of currents impinging 
on one node by the same bond will describe a scalar, a spinor 
or a vector field. Since they need only one type of current, 
scalar waves correspond to the simplest situation. For the sake 
of illustration, they will be the focus of this paper. In addition 
to the propagating currents, a current is attached to each node 
in order to describe an inhomogeneous medium. This current does 
not propagate but participates in the scattering process which 
involves the propagating currents. We shall demonstrate that 
a few elementary properties and symmetries of the scatterers 
like time reversal symmetry, reciprocity and isotropy lead to 
the classical wave, Klein-Gordon or Schr\"{o}dinger equations. 
At that stage, we find that these discrete equations are not 
yet completely general because the mass, the velocity or the 
potential remain correlated. We show that it is possible to remove 
this limitation by attaching a second current to each node.%

The paper is organized as follows. In section~\ref{cf}, we introduce 
the currents which propagate on the mesh of a Cartesian lattice. 
Scatterers are described by scattering matrices located on each 
node of the lattice. The field is then defined as a function 
of incident currents. In order for the field to obey a discretized 
wave equation, the currents must be eliminated from its time 
evolution. We show in section~\ref{cl} how this condition strongly compels 
the form of the scattering matrices. Time-reversal symmetry, 
reciprocity and isotropy of the scatterers are introduced in 
sections~\ref{tr}, \ref{re} and~\ref{is} respectively. 
The resulting discrete wave equations are discussed in section~\ref{we}. 
This leads us to attach a second current to each node 
as described in section~\ref{tn}. We conclude 
in section~\ref{co} by discussing possible extensions of this work.%


\section{Currents and fields}~\label{cf}
\subsection{Definition of the currents}~\label{dc}
\begin{figure}[htb]
\begin{center}
\epsffile{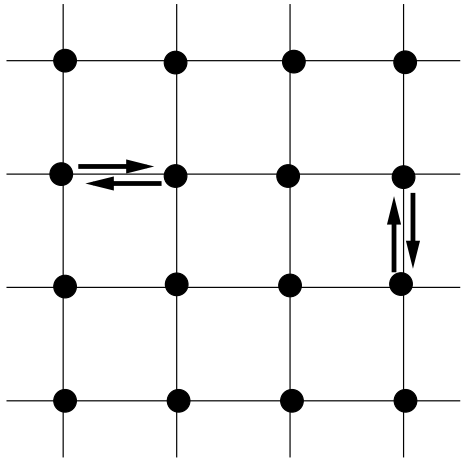}
\caption{Currents propagating along the bonds of a two-dimensional 
Euclidean lattice.}~\label{fig:cartesien}
\end{center}
\end{figure}

A current is a real or complex number, which propagates along 
a bond of a $d$-dimensional Cartesian lattice. For scalar waves, 
it is sufficient to consider that each lattice bond carries two 
currents propagating in opposite directions 
(Figure \ref{fig:cartesien}).\footnote{The
results described in this paper are valid for any $d$-dimensional space.
However, for convenience, all figures will represent a two-dimensional
Cartesian lattice.} At any time $t$, the system is completely 
defined by the values 
of all currents in the lattice. We assume that time and space 
variables are discrete. In one time step $\tau$, 
a current propagates between two nodes of a given bond. All currents 
are synchronized. In other words, the currents are simultaneously 
outgoing from the nodes at some time $t$ and become incident currents 
on neighbor nodes at time $t+\tau$. 

\begin{figure}[htb]
\begin{center}
\epsffile{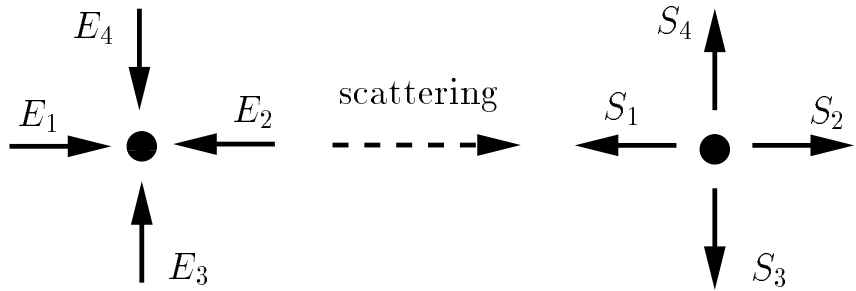}
\caption{Scattering process.}~\label{fig:scattering}
\end{center}
\end{figure}

Each node of the lattice is a scatterer. The scatterers are identical 
in a periodic system or are distinct, and randomly chosen in 
a disordered quenched system. Each scatterer is described by 
a $2d\times 2d$ matrix $S$ which transforms the $2d$ incident currents $E_l$, 
at any time $t$, in $2d$ outgoing currents $S_k$ at the same time 
(Figure \ref{fig:scattering}) according to

\[
S_k = \sum_{l=1}^{2d} s_{kl} E_l \esp k=1,\ldots,2d,
\]
where the $s_{kl}$ are the matrix elements of the matrix $S$.

It turns out that the model is not sufficient to describe wave 
propagation in an inhomogeneous medium like a material in which 
the index of refraction is a function of the position. As shown 
later, this problem is solved by attaching to each node an additional 
current that will be called throughout this paper the node current. 
For notational convenience in sums over the currents, we shall 
write the node currents $E_{2d+1}$ and $S_{2d+1}$. However, we emphasize 
that the subscript $2d+1$ is not ascribed to a spatial dimension 
in contrast with the other indices ranging from $1$ to $2d$. The node 
current participates in the same scattering process as the other currents 
but does not propagate to neighbor nodes. Instead, the outgoing current 
$S_{2d+1}$ becomes the incident current $E_{2d+1}$ on the same node at the 
next time step. In other words, the propagation process for the node 
current reads $E_{2d+1}(t+\tau)=S_{2d+1}(t)$. We can imagine the node 
current as propagating in one time step along a loop attached to the node. 
This loop is equivalent to the ``permittivity stub'' introduced in the 
TLM method to describe inhomogeneous electromagnetic structures\cite{johns74b}.

\begin{figure}[htb]
\begin{center}
\epsffile{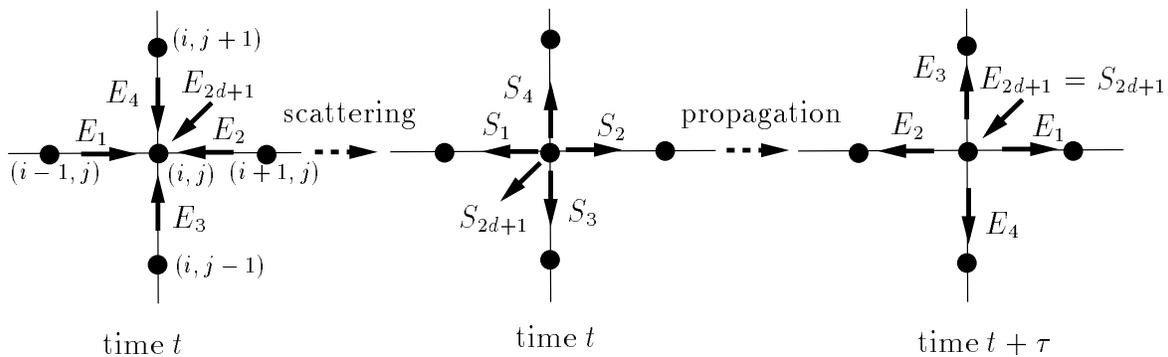}
\caption{Time evolution of the currents.}~\label{fig:def_evolution}
\end{center}
\end{figure}

In conclusion, the general scheme of the time evolution of the 
currents is suitably represented in Fig.\ \ref{fig:def_evolution}. 

We write

\begin{equation}
S_{k} = \sum_{l=1}^{2d+1} s_{kl} E_{l}, \esp k=1,\ldots,2d+1,
\label{def:diffusion}
\end{equation} 
where the $s_{kl}$ are the matrix elements of the $(2d+1)\times (2d+1)$ 
scattering matrix $S$. This matrix is a function of the position in an 
inhomogeneous medium.

\subsection{Definition of the field}~\label{df}
The previous definition of the currents includes the three essential 
features of the model~: the principle of action-by-proximity, 
which is described by the current propagation between neighbor 
nodes, the scattering process in which the nodes act like secondary 
sources of ``spherical'' wavelets and the linearity of this process. 
These features can be viewed as a discretized realization of 
Huygens' principle. However, additional ingredients and choices 
are needed to achieve the description of the model. For example, 
the form of the field propagation equation will strongly  depend 
on the choice of the scattering matrices. Among all the questions 
to be answered in our approach, the first one is how to define 
the field.

The field is supposed to obey a discretized linear wave equation 
which depends on the physical problem we are interested in~: the 
scalar wave equation for an acoustic wave, the Klein-Gordon equation 
for a zero-spin massive particle, etc... At time $t$, the field 
is defined as a real or complex quantity $\psi(i_{1},i_{2},\ldots,i_{d},t)$ 
on each node of the lattice, where $i_{1},i_{2},\ldots,i_{d}$, 
are the discrete indices which locate the node position 
$x_{k}=i_{k} l, k=1,\ldots,d$, as a function of the mesh parameter 
$l$. For briefness, we note $\vect{r}$ the set of indices 
$i_{1},i_{2},\ldots,i_{d}$, of an arbitrary node. There is no reason for 
the field to be identified with one of the currents previously defined. 
Actually, we just postulate that the value of the field $\psi(\vect{r},t)$ 
is a function of the incident and outgoing currents at node $\vect{r}$
and at time $t$. Therefore, by taking account of linearity, the most general 
definition reads

\[
\psi(\vect{r},t) = \sum_{k=1}^{2d+1} \left[\lambda_{k}^{\prime}(\vect{r}) 
E_{k}(\vect{r},t)+\lambda_{k}^{\prime\prime}(\vect{r}) 
S_{k}(\vect{r},t)\right],
\]
where $\lambda_{k}^{\prime}$ and $\lambda_{k}^{\prime\prime}$ are complex 
coefficients to be determined. 

The field is a linear superposition of the currents, which vanishes 
when all currents vanish. By noticing that, due to the scattering 
process \refp{def:diffusion}, the outgoing currents $S_{k}(\vect{r},t)$
can be expressed as linear combinations of the incident currents 
$E_{k}(\vect{r},t)$ at the same time $t$, it is sufficient to define the 
field solely as a function of the incident currents

\begin{equation}
\psi(\vect{r},t) = \sum_{k=1}^{2d+1} \lambda_{k}(\vect{r}) E_{k}(\vect{r},t).
\label{def:field}
\end{equation}
It is necessary to point out that the $\lambda_{k}$ are functions of the 
node position. This property is required to describe an inhomogeneous system.


\section{Closure of the wave equation}~\label{cl}
The general form of the discretized wave equations we 
are looking for  can be written~:

\begin{equation}
\psi(\vect{r},t+\tau) = f \left( 
\psi(\vect{r}^{\prime},t'),\psi(\vect{r}^{\prime\prime},t''),
\psi(\vect{r}^{\prime\prime\prime},t'''),\ldots \right),
\label{eq:closure}
\end{equation}
where the field at node $\vect{r}$ and time $t+\tau$ is a function 
of the field on the same node and/or on nodes at previous times 
$t',t''$, etc$\ldots$ Such an equation is straightforwardly derived 
from the discretization of the corresponding continuous equation. 
Equation \refp{eq:closure} is a closed 
equation, which means that $f$ is a function where no currents appear 
explicitely (so that \refp{eq:closure} involves fields exclusively), 
and that the number of terms in the r.h.s. of \refp{eq:closure} must be 
finite. Usually, and at the lowest order of the discretization procedure, 
the involved fields are fields on the same node 
$\vect{r}=\left\{i_{1},i_{2},\ldots,i_{d}\right\}$ and on the neighbor 
nodes $i_{1}\pm 1,i_{2}\pm 1,\ldots,i_{d}\pm 1$, at times $t$ and $t-\tau$. 
In the remainder of this section we derive, by use of our model, 
a closed equation of the form of equation \refp{eq:closure}. Also, we show 
that imposing the closure conditions strongly constrains the 
form of the scattering matrix.

\begin{figure}[htb]
\begin{center}
\epsffile{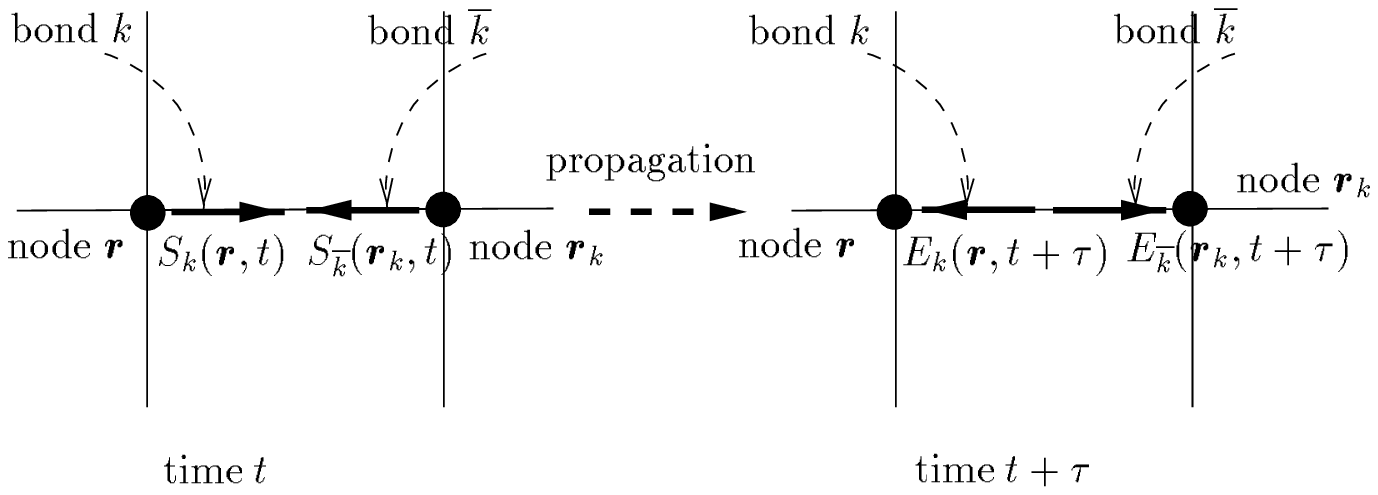}
\caption{Notation for the bond linking the central node $\vect{r}$ to 
one of its neighbor nodes $\vect{r}_{k}$ and the propagating currents 
defined on this bond.}~\label{fig:def_notbond}
\end{center}
\end{figure}

For this purpose, we start to derive an equation of the type 
of Eq.\ \refp{eq:closure} from the evolution rules summarized in 
Fig.\ \ref{fig:def_evolution} of the model. 
First of all, let us introduce some notations. 
In a $d$-dimensional Euclidean space, node $\vect{r}$ is linked to the 
neighbor nodes by $2d$ bonds.  We note $\vect{r}_{k}$ the set of indices 
of the neighbor node which is linked to node $\vect{r}$ by the bond $k$, along 
which the currents $E_{k}(\vect{r})$ or $S_{k}(\vect{r})$ propagate 
(Figure \ref{fig:def_notbond}). Moreover, bond $k$ of node $\vect{r}$ is 
named bond $\overline{k}$ for node $\vect{r}_{k}$, which simply means, 
for example, that bond $2$ of the central node in Figure \ref{fig:def_notbond}
corresponds to bond $1$ of the right side node $(k=2,\overline{k}=1)$. \
This notation immediatly leads to 
$E_{\overline{k}}(\vect{r}_{k},t+\tau) = S_{k}(\vect{r},t)$ and 
$E_{k}(\vect{r},t+\tau) = S_{\overline{k}}(\vect{r}_{k},t)$. 
Note that for the node current $(k=2d+1)$, we have $\vect{r}_{2d+1}=\vect{r}$
and $\overline{k}=k$. 

The first step in attempting to derive an equation like \refp{eq:closure}, 
is to write down the definition of the field $\psi(\vect{r},t+\tau)$ on a 
node $\vect{r}$ at time $t+\tau$

\begin{eqnarray}
\psi(\vect{r},t+\tau) & = & \sum_{k=1}^{2d+1} \lambda_{k}(\vect{r}) 
E_{k}(\vect{r},t+\tau), \nonumber \\
& = & \sum_{k=1}^{2d+1} \lambda_{k}(\vect{r}) S_{\overline{k}}(\vect{r}_{k},t),
\nonumber \\
& = & \sum_{k=1}^{2d+1} \lambda_{k}(\vect{r})\left[ \sum_{l=1}^{2d+1} 
s_{\overline{k} l}(\vect{r}_{k}) E_{l}(\vect{r}_{k},t) \right],
\label{eq:field_out1}
\end{eqnarray} 
where the propagation and the scattering rules have been used.

Thus, the field $\psi(\vect{r},t+\tau)$ at time $t+\tau$ is expressed 
in terms of the incident currents at time $t$ on the same node $\vect{r}$ 
and on its neighbor $\vect{r}_{k}$. One realizes immediately that using 
the same rules to express the incident currents at time $t$ as a function of 
the incident currents at time $t-\tau$ and iterating the procedure, will 
give incident currents at times $t-\tau,t-2\tau,\ldots$, on nodes as distant 
from node $\vect{r}$ as desired. Such a process is not bounded and seems 
to prevent us from writing a closed equation for the field $\psi$. This 
statement is true except for some particular scattering matrices of the 
type we describe below. For this purpose, we shall rewrite the scattering 
equation \refp{def:diffusion} in order to exhibit the field $\psi$~:

\begin{equation}
S_{k} = \sum_{k=1}^{2d+1} s_{kl}E_{l} = 
\displaystyle{\frac{s_{km}}{\lambda_{m}}} \psi + \sum_{l=1}^{2d+1} 
(s_{kl} - \lambda_{l}\displaystyle{\frac{s_{km}}{\lambda_{m}}}) E_{l},
\label{eq:scatt1}
\end{equation}
where $m$ is any of the  $2d+1$ current indices. Suppose that the following 
condition is satisfied~: 

\begin{equation}
\displaystyle{\frac{s_{kl}}{\lambda_{l}}} = \mbox{constant} \equiv \rho_{k},
\esp \forall l,
\label{eq:scatt2}
\end{equation}
then, equation \refp{eq:scatt1} becomes~:

\[
S_{k}= \rho_{k} \psi,
\]
and substitution in equation \refp{eq:field_out1} leads directly to~:

\begin{equation}
\psi(\vect{r},t+\tau) = \sum_{k=1}^{2d+1} \lambda_{k}(\vect{r})
\rho_{\overline{k}}(\vect{r}_{k}) \psi(\vect{r}_{k},t),
\label{eq:field_out2}
\end{equation}
which is a closed equation of the type we are looking for. It turns out 
that condition \refp{eq:scatt2} is too restrictive and can be replaced by

\begin{equation}
\displaystyle{\frac{s_{kl}}{\lambda_{l}}} = \mbox{constant} \equiv \rho_{k},
\esp \forall l \neq k,
\label{eq:scatt3}
\end{equation}
Then, equation \refp{eq:scatt1} becomes 

\begin{equation}
S_{k} = \rho_{k}\psi - \mu_{k} E_{k},
\label{eq:scatt4}
\end{equation}
where $\mu_{k}=\rho_{k}\lambda_{k} - s_{kk}$.

\begin{figure}[htb]
\begin{center}
\epsffile{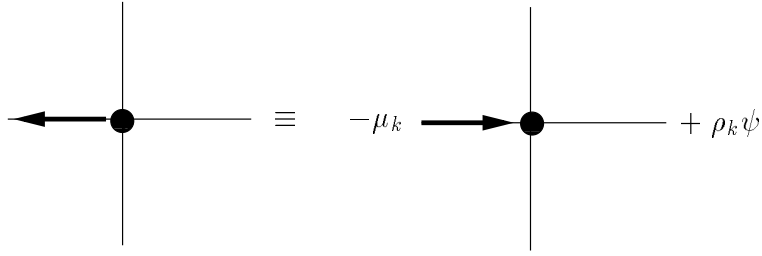}
\caption{The scattering process depicted according to equation 
\refp{eq:scatt4}.}~\label{fig:smatrix}
\end{center}
\end{figure}

According to \refp{eq:scatt4}, the outgoing current on one bond is not only 
a function of the field $\psi$ but also depends on the incident current on 
the same bond as sketched Figure \ref{fig:smatrix} 

\begin{figure}[htb]
\epsffile{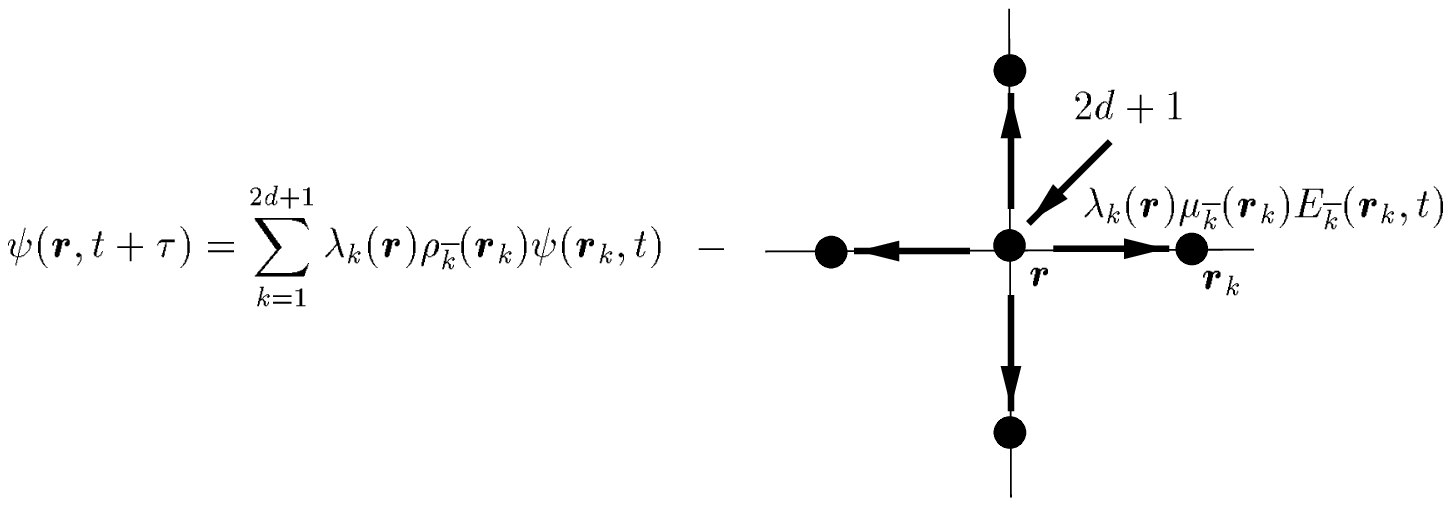}
\caption{Sketch of equation~\refp{eq:field_out3}.}~\label{fig:psi_1}
\end{figure}

It seems at first sight that we went 
backward in writing condition \refp{eq:scatt3} instead of \refp{eq:scatt2}, 
since a current appears in equation \refp{eq:scatt4} in addition to the 
field $\psi$. Actually, substitution of equation \refp{eq:scatt4} in equation 
\refp{eq:field_out1} gives

\begin{equation}
\psi(\vect{r},t+\tau) = \sum_{k=1}^{2d+1} \lambda_{k}(\vect{r})
\rho_{\overline{k}}(\vect{r}_{k}) \psi(\vect{r}_{k},t) - \sum_{k=1}^{2d+1}
\lambda_{k}(\vect{r}) \mu_{\overline{k}}(\vect{r}_{k}) 
E_{\overline{k}}(\vect{r}_{k},t),
\label{eq:field_out3}
\end{equation}  
which is represented schematically on Figure \ref{fig:psi_1}.

\begin{figure}[htb]
\begin{center}
\epsffile{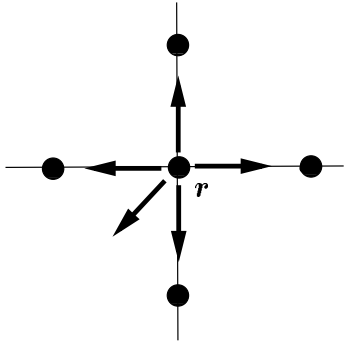}
\caption{The outgoing currents leaving node $\vect{r}$ at time 
$t-\tau$, which give after propagation the input currents at time $t$ 
in equation \refp{eq:field_out3}.}~\label{fig:psi_2}
\end{center}
\end{figure}

It is obvious from Figure 
\ref{fig:psi_1} that the $2d+1$ incident currents 
$E_{\overline{k}}(\vect{r}_{k},t)$
at time $t$ (bold arrows) are deduced, according to the propagation rules, 
from the currents $S_{k}(\vect{r},t-\tau)$ leaving node 
$\vect{r}$ at time $t-\tau$, as depicted in Figure \ref{fig:psi_2}.

Thus, the additional currents due to the looser condition \refp{eq:scatt3} 
eventually involve only the initial node $\vect{r}$ instead of propagating 
new terms to remote nodes. Using again equation \refp{eq:scatt4}, 
equation \refp{eq:field_out3} becomes

\begin{eqnarray}
\lefteqn{\psi(\vect{r},t+\tau) = \sum_{k=1}^{2d+1} \lambda_{k}(\vect{r})
\rho_{\overline{k}}(\vect{r}_{k}) \psi(\vect{r}_{k},t)} \nonumber \\
&& - \left[ \sum_{k=1}^{2d+1} \lambda_{k}(\vect{r})\mu_{\overline{k}}(\vect{r}_{k})
\rho_{k}(\vect{r}) \right] \psi(\vect{r},t-\tau) + 
\sum_{k=1}^{2d+1} \lambda_{k}(\vect{r})\mu_{\overline{k}}(\vect{r}_{k}) 
\mu_{k}(\vect{r})E_{k}(\vect{r},t-\tau),
\label{eq:field_out4}
\end{eqnarray}
The last term is a linear superposition of $2d+1$ incident currents upon node
$\vect{r}$ at time $t-\tau$. In order for equation \refp{eq:field_out4} to be 
a closed field equation without current terms, it is sufficient for this 
superposition of currents to be proportional to 
$\psi(\vect{r},t-\tau) = \sum_{k=1}^{2d+1} \lambda_{k}(\vect{r}) 
E_{k}(\vect{r},t-\tau)$. Therefore, we must set%

\begin{equation}
\mu_{\overline{k}}(\vect{r}_{k}) \mu_{k}(\vect{r}) = \mu^2(\vect{r}), \esp
\forall k=1,\ldots,2d+1,
\label{eq:condsuf1}
\end{equation}
where we denote $\mu^2(\vect{r})$ a constant which is a priori a function of 
$\vect{r}$. In fact, it is simple to prove that $\mu^2(\vect{r})$ does not 
depend on $\vect{r}$. To prove this we begin by recognizing that equation 
\refp{eq:condsuf1} can be written for one of the neighbor node $\vect{r}_{k}$

\[
\mu_{\overline{l}}(\vect{r}_{kl})\mu_{l}(\vect{r}_{k}) = \mu^2(\vect{r}_{k}),
\esp \forall l=1,\ldots,2d+1.
\]
Now $\vect{r}_{k}$ plays the role of the central node $\vect{r}$ of 
equation \refp{eq:condsuf1} and $\vect{r}_{kl}$ denotes its neighbor nodes. 
Among the neighbor nodes $\vect{r}_{kl}$, the initial node $\vect{r}$ 
$(\overline{l}=k)$ is of particular interest and the last equation reads

\[
\mu_{k}(\vect{r})\mu_{\overline{k}}(\vect{r}_{k}) = \mu^2(\vect{r}_{k}).
\]       
Comparison with equation \refp{eq:condsuf1} leads to the required identity

\[
\mu^2(\vect{r}_{k}) = \mu^2(\vect{r}).
\]
Since the node $\vect{r}$ is arbitrary, equation \refp{eq:condsuf1} can be 
recast in a simpler form valid for any node

\begin{equation}
\mu_{\overline{k}}(\vect{r}_{k})\mu_{k}(\vect{r}) = \mu^2, \esp \forall k,
\label{eq:condsuf3}
\end{equation}
where now the constant $\mu^2$ is independent of the node's position.

One notices that for $k=2d+1$, $\vect{r}_{k}=\vect{r}$, $\overline{k}=k$, 
and $\mu_{\overline{k}}(\vect{r}_{k}) = \mu_{k}(\vect{r})$. Therefore

\begin{equation}
\mu_{2d+1}(\vect{r}) = \epsilon_{2d+1}(\vect{r}) \mu,
\label{eq:condsuf4}
\end{equation}
where $\epsilon_{2d+1}(\vect{r})=\pm 1$.

Finally, utilizing equation \refp{eq:condsuf3}, equation \refp{eq:field_out4} 
reads

\begin{equation}
\psi(\vect{r},t+\tau) = \sum_{k=1}^{2d+1} \lambda_{k}(\vect{r})
\rho_{\overline{k}}(\vect{r}_{k}) \psi(\vect{r}_{k},t) + 
\mu^2 \left[ 1 - \sum_{k=1}^{2d+1} 
\displaystyle{\frac{\lambda_{k}(\vect{r})\rho_{k}(\vect{r})}{\mu_{k}(\vect{r})}} \right] \psi(\vect{r},t-\tau).
\label{eq:field_out5}  
\end{equation}
Equation \refp{eq:field_out5} is a closed field equation in which currents do 
not appear. As the field $\psi$ is exhibited at times $t+\tau$, $t$  and 
$t-\tau$, this new equation is an improvement over equation 
\refp{eq:field_out2} in the sense that a discrete second time derivative 
$\psi(t+\tau) -2\psi(t)+\psi(t-\tau)$ can be obtained as required in, 
for example, the discrete time-dependent wave equation.
Making use of condition \refp{eq:scatt3}, the general matrix element reads

\begin{equation}
s_{kl} = \rho_{k}\lambda_{l} - \mu_{k}\delta_{kl},
\label{eq:scatt5}
\end{equation}
where $\mu_{k}=\rho_{k}\lambda_{k} - s_{kk}$ and $\delta_{kl}$ is the 
Kronecker symbol.
Instead of $(2d+1)^2$ elements, the scattering matrix depends now on the 
$3(2d+1)$ complex coefficients $\rho_{k}$, $\lambda_{k}$ and $\mu_{k}$. 
These coefficients vary independently from one node to another 
in an inhomogeneous system except for the coefficients $\mu_{k}$ which obey 
condition \refp{eq:condsuf3} relating neighbor nodes. Furthermore, 
$\mu_{2d+1}=\pm\mu$, according to equation \refp{eq:condsuf4}.

In summary, equation \refp{eq:field_out5} is the general equation which 
describes the time evolution of the field $\psi$. This closed equation has 
been derived from the evolution rules of the currents, the definition 
of $\psi$ \refp{def:field} and from conditions \refp{eq:scatt3} 
and \refp{eq:condsuf3}. Let us discuss the status of these conditions. 
Condition \refp{eq:scatt3} cannot be avoided if we look for a closed 
equation in $\psi$, since a looser condition would make terms appear on 
all nodes of the system. Meanwhile, condition \refp{eq:condsuf3}, which 
allows one to replace the currents in the last term of equation 
\refp{eq:field_out4} by the field $\psi(\vect{r},t-\tau)$, is sufficient but 
not necessary. Indeed, suppose we consider that these incident currents 
defined on node $\vect{r}$ at time $t-\tau$ are the result of propagation 
and scattering of currents at previous times $t-2\tau$ and $t-3\tau$ on 
the neighbor nodes and on node $\vect{r}$. Reproducing the same calculation 
steps as those which lead to equation \refp{eq:field_out4}, the last term 
may be replaced on the one hand by new terms involving the field $\psi$
on node $\vect{r}$ at time $t-3\tau$ and on its neighbors $\vect{r}_{k}$ at 
time $t-2\tau$ respectively, and on the other hand by a new linear 
superposition of incident currents on node $\vect{r}$ at time 
$t-3\tau$. Therefore, it is possible to recover a new closure condition 
(analogous to equation \refp{eq:condsuf3}) by taking this new linear 
superposition of incident currents to be proportional to 
$\psi(\vect{r},t-3\tau)$. Additionaly, the same procedure can be iterated 
and similar superpositions appear at times $t-5\tau$, $t-7\tau$, etc$\ldots$
So that the resulting  field equation can be closed at any step of the 
iteration procedure. The net result is to open the possibility of describing 
high order time derivatives of the field $\psi$. We shall not consider these 
possibilities further and restrict ourselves with equation 
\refp{eq:field_out5} which is sufficient to obtain the 
second order time derivative of $\psi$.

 The description of the model is far from being achieved. To 
make further progress, additional properties of the scatterers 
are needed. The most basic properties to consider are suitable 
symmetries of the scattering process which will restrict again 
the number of independent parameters and make precise the physical 
content of equation \refp{eq:field_out5}. We shall successively introduce some 
of the most natural symmetries we can think of, namely time-reversal 
symmetry, reciprocity and isotropy. It is worthwile to point 
out that such symmetries, like the results obtained so far, will 
only involve the discrete geometry of the Cartesian lattice and 
assume neither any topological space nor any Euclidean or Riemannian 
structure. Nevertheless, we shall need the definition of underlying 
manifolds provided with such structures when taking later the 
continuous limit of the discrete wave equation obeyed by the 
field $\psi$.


\section{Time-reversal symmetry}~\label{tr}
In this section, we introduce the time-reversal symmetry by formulating 
its natural action on both currents and fields.

\subsection{Currents}~\label{ic}
\begin{figure}[htb]
\begin{center}
\epsffile{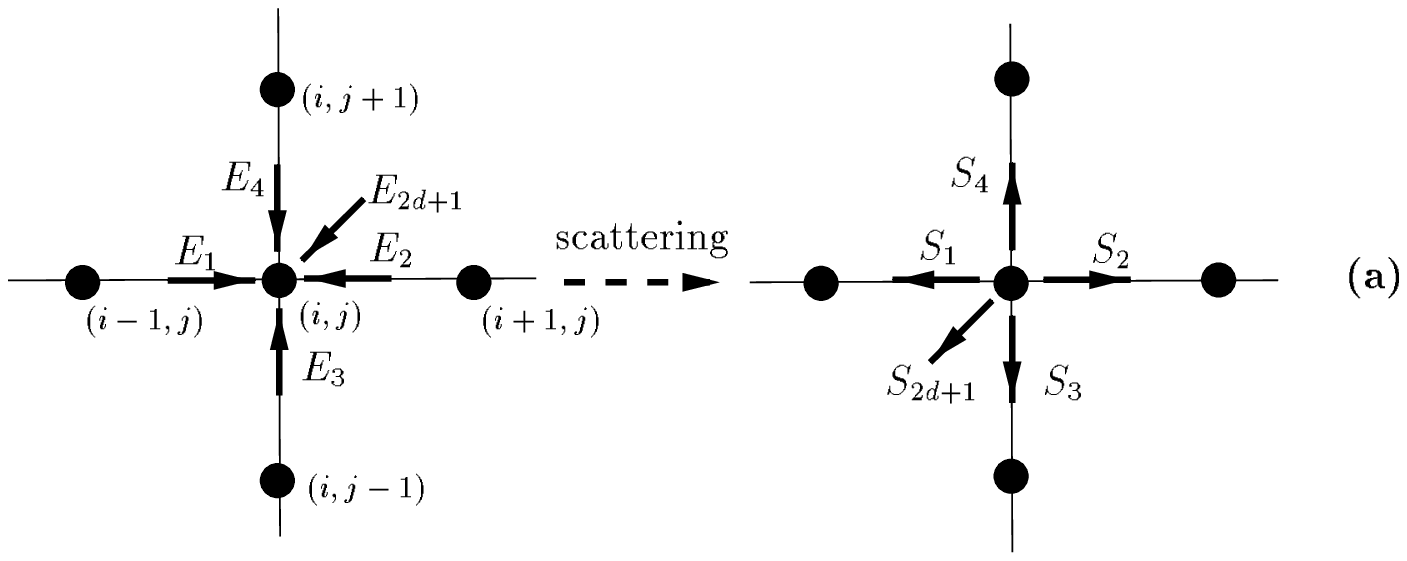}
\epsffile{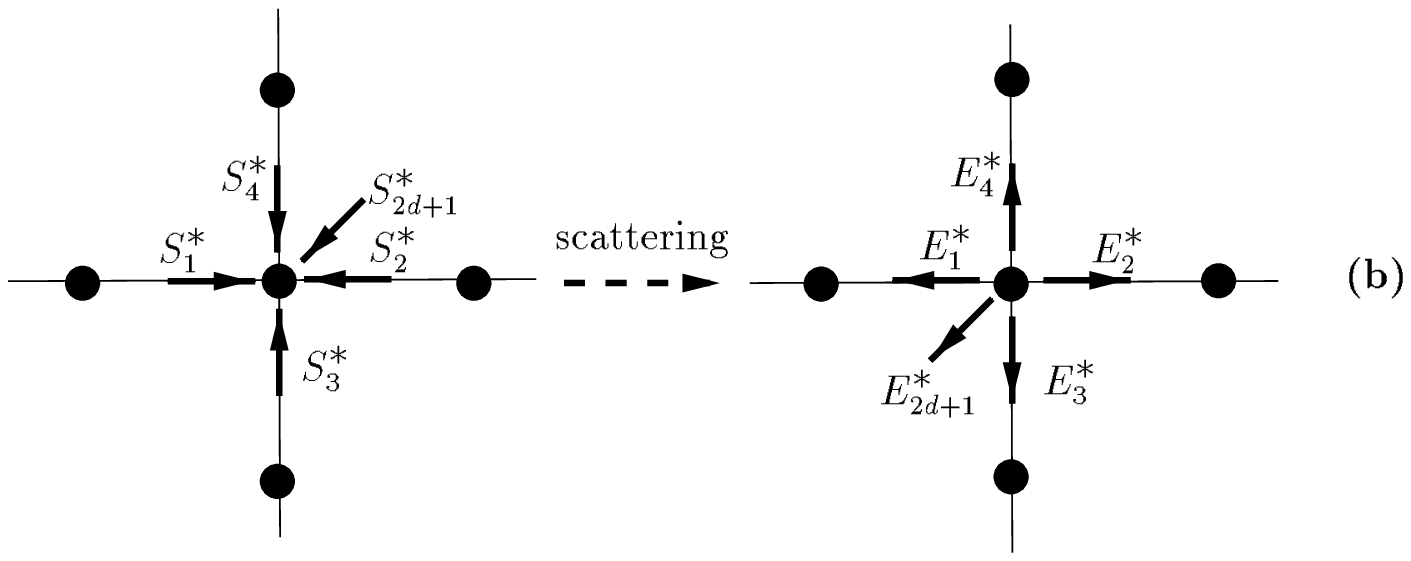}
\caption{Processes involved in the time-reversal symmetry. 
{\bf (a)} forward process. {\bf (b)} 
time-reversed process.}~\label{fig:it_scalar}
\end{center}
\end{figure}

Time-reversal is naturally completed by reversing the 
direction of the current arrows in such a way that the currents 
propagate and are scattered backward in time. They will obey 
time-reversal symmetry only if they describe the past states 
of the system in reverse order. It is then obvious that the scattering 
process must be reversible. Therefore, we state that the forward 
scattering process  depicted in Figure \ref{fig:it_scalar}~{\bf (a)} 
implies the existence of the reverse process  depicted in Figure 
\ref{fig:it_scalar}~{\bf (b)}. Since the currents 
are, in general, complex quantities, complex conjugate currents 
are involved in the time-reversed process. If the currents are 
real, this will be equivalent to reversing the original currents.

Using matrix notation,  time-reversal invariance reads

\[
\left(S_{k}\right) = S \left(E_{k}\right) \Longrightarrow 
\left(E_{k}^{\ast}\right) = S \left(S_{k}^{\ast}\right),
\]
where $\left(S_{k}\right)$ and $\left(E_{k}\right)$ are the column vectors 
of outgoing and incident currents respectively.

Then, taking the complex conjugate 
$\left(E_{k}\right)=S^{\ast}\left(S_{k}\right)$, one obtains

\[
S S^{\ast} = S^{\ast} S = I_{2d+1},
\]
where $I_{2d+1}$ is the $(2d+1)\times (2d+1)$ unit matrix.

This condition and the general form \refp{eq:scatt5} for the scattering 
matrix elements leads to the following conditions
 
\begin{equation}
\left\{
\begin{array}{ccccccl}
\displaystyle{\frac{\mu_{k}\rho_{k}^{\ast}}{\rho_{k}}} & = & \mbox{constant}
& = & e^{i\gamma_{1}}, && \\
\\
\displaystyle{\frac{\mu_{k}\lambda_{k}^{\ast}}{\lambda_{k}}} & = & 
\mbox{constant} & = & e^{i\gamma_{2}}, && \esp \forall k=1,\ldots,2d+1,\\
\\
\module{\mu_{k}} & = & 1, &&  
\end{array}
\right.
\label{eq:it_currents1}
\end{equation}
and

\begin{equation}
\Lambda_{1} \equiv \sum_{k=1}^{2d+1} \lambda_{k}\rho_{k}^{\ast}
= e^{i\gamma_{1}} + e^{-i\gamma_{2}}.
\label{eq:it_currents2}
\end{equation}
Again, let us point out that all the quantities appearing in 
these relations are functions of the node $\vect{r}$. This statement holds 
in particular for the constants $C_{1}=e^{i\gamma_{1}}$ and 
$C_{2}=e^{i\gamma_{2}}$.

\subsection{Time-reversal symmetry of the field}~\label{if}
Equation \refp{eq:field_out5} which describes the time evolution 
of the field can be viewed as some general relation which expresses 
the field $\psi$ at time $t+\tau$ as a function of $\psi$ at times $t$ and 
$t-\tau$

\begin{equation}
\psi(\vect{r},t+\tau) = f \left[ \psi(\vect{r}_{k},t),\psi(\vect{r},t-\tau)
\right].
\label{eq:field_out6}
\end{equation}
In general, $\psi$ is  complex valued function so that equation 
\refp{eq:field_out6} obeys time-reversal symmetry if the reverse-time 
equation is  verified by the complex conjugate field $\psi^{\ast}$, i.e.,

\begin{equation}
\psi^{\ast}(\vect{r},t-\tau) = f \left[ \psi^{\ast}(\vect{r}_{k},t),
\psi^{\ast}(\vect{r},t+\tau)\right].
\label{eq:field_it1}
\end{equation}
Therefore, our task is to check that the time-reversal symmetry translated 
on currents, which was considered in the previous section, enables 
us to obtain equation \refp{eq:field_it1} from equation \refp{eq:field_out6}.
For this purpose, we first substitute the definition \refp{def:field} of 
the field $\psi$ in \refp{eq:field_out6}

\begin{equation}
\sum_{k=1}^{2d+1} \lambda_{k}(\vect{r}) E_{k}(\vect{r},t+\tau) = 
f \left[ \sum_{l=1}^{2d+1} \lambda_{l}(\vect{r}_{k}) E_{l}(\vect{r}_{k},t),
\sum_{k=1}^{2d+1} \lambda_{k}(\vect{r}) E_{k}(\vect{r},t-\tau) \right]
\label{eq:field_it2}
\end{equation}
Then, Figure \ref{fig:it_scalar} tells us that reversing the direction 
of current propagation and substituting the complex conjugate currents 
$S^{\ast}_{k}$ for $E_{k}$ gives the current values of the 
reverse-time system. In other words, the time-reversal symmetry 
applied to the currents, appearing in equation \refp{eq:field_it2}, leads to

\begin{equation}
\sum_{k=1}^{2d+1} \lambda_{k}(\vect{r}) S^{\ast}(\vect{r},t-\tau) = 
f \left[ \sum_{l=1}^{2d+1} \lambda_{l}(\vect{r}_{k}) 
S^{\ast}_{l}(\vect{r}_{k},t), \sum_{k=1}^{2d+1} \lambda_{k}(\vect{r}) 
S^{\ast}(\vect{r},t+\tau) \right].
\label{eq:field_it3}
\end{equation}
This equation is similar to equation \refp{eq:field_it1}. However, 
it involves the quantities $\sum_{k=1}^{2d+1} \lambda_{k} S^{\ast}_{k}$
instead of $\psi^{\ast} = \sum_{k=1}^{2d+1}\lambda^{\ast}_{k} E^{\ast}_{k}$. 
Therefore, to demonstrate the time-reversal invariance of the 
field $\psi$, it is necessary to show that these two quantities are identical. 
Actually, they just need to be proportional to each other since 
the function $f$ is linear. Using the scattering formula 
\refp{eq:scatt4}, we can write

\[
\sum_{k=1}^{2d+1} \lambda_{k} S^{\ast}_{k} = \Lambda_{1}\psi^{\ast}
- \sum_{k=1}^{2d+1} \lambda_{k}\mu^{\ast}_{k} E^{\ast}_{k}.
\]
Since $\lambda_{k}\mu^{\ast}_{k} = C^{\ast}_{2}\lambda^{\ast}_{k}$ according 
to the second equation of \refp{eq:it_currents1}, one obtains

\[
\sum_{k=1}^{2d+1} \lambda_{k} S^{\ast}_{k} = (\Lambda_{1} - C^{\ast}_{2})
\psi^{\ast} = C_{1}\psi^{\ast},
\] 
where we have used equation \refp{eq:it_currents2}. However, we have 
previously pointed out that $C_{1}$ is a function of the node $\vect{r}$. 
Therefore, substituting $C_{1}\psi^{\ast}$ for $\sum_{k=1}^{2d+1} \lambda_{k}
S^{\ast}_{k}$ in equation \refp{eq:field_it3} leads to

\[
C_{1}(\vect{r})\psi^{\ast}(\vect{r},t-\tau) = f \left[
C_{1}(\vect{r}_{k})\psi^{\ast}(\vect{r}_{k},t),
C_{1}(\vect{r})\psi^{\ast}(\vect{r},t+\tau) \right].
\]
This equation will be equivalent to equation \refp{eq:field_it1} if we 
impose the condition

\[
C_{1}(\vect{r}_{k}) = C_{1}(\vect{r}), \esp \forall k.
\]
Thus, the constant $C_{1}$ must be the same on the node $\vect{r}$ 
and on its neighbors $\vect{r}_{k}$, for any node $\vect{r}$ in the system. 
One must conclude that $C_{1}$ is a constant over the whole lattice. 
Therefore,  the first equation of \refp{eq:it_currents1} becomes

\begin{equation}
\displaystyle{\frac{\mu_{k}(\vect{r})\rho^{\ast}_{k}(\vect{r})}
{\rho_{k}(\vect{r})}} = e^{i\gamma_{1}}, \esp \forall \vect{r},k.
\label{eq:it_currents3}
\end{equation}

\subsection{Form of the time-dependent equation}~\label{fe}
We now proceed to show that the results obtained so far have 
already strong consequences on the form of the evolution equation 
of $\psi$. First, the coefficients of $\psi(\vect{r},t-\tau)$ and 
$\psi(\vect{r}_{k},t)$ in equation \refp{eq:field_out5} can be recast in 
the following form, with the help of identities \refp{eq:it_currents1}, 
\refp{eq:it_currents2} and \refp{eq:it_currents3}

\[
\mu^2 \left[1 - \sum_{k=1}^{2d+1} 
\displaystyle{
\frac{\lambda_{k}(\vect{r})\rho_{k}(\vect{r})}{\mu_{k}(\vect{r})}} \right]
= -\displaystyle{\frac{\mu^2}{C_{1}C_{2}(\vect{r})}},
\]
and 

\[
\lambda_{k}(\vect{r})\rho_{\overline{k}}(\vect{r}_{k}) = 
\epsilon_{k}(\vect{r}) \module{\lambda_{k}(\vect{r})}
\module{\rho_{\overline{k}}(\vect{r}_{k})}\left[ \displaystyle{\frac{\mu^2}
{C_{1}C_{2}(\vect{r})}}\right]^{1/2} ,
\]
where the modulus and phases of $\lambda$ and  $\rho$ have been separated in 
the above expression. The sign $\epsilon_{k}(\vect{r})=\pm 1$ depends on the 
choice made in taking the square root. Plugging these two results into 
\refp{eq:field_out5} leads to the time-reversal invariant equation of the 
field

\begin{eqnarray}
\psi(\vect{r},t+\tau) + 
\displaystyle{\frac{\mu^2}{C_{1}C_{2}(\vect{r})}}
\psi(\vect{r},t-\tau) & = & \sum_{k=1}^{2d+1} \lambda_{k}(\vect{r})
\rho_{\overline{k}}(\vect{r}_{k}) \psi(\vect{r}_{k},t), \nonumber \\
\nonumber \\
& = & \left[ \displaystyle{\frac{\mu^2}{C_{1}C_{2}(\vect{r})}}\right]^{1/2}
\sum_{k=1}^{2d+1} \epsilon_{k}(\vect{r}) \module{\lambda_{k}(\vect{r})}
\module{\rho_{\overline{k}}(\vect{r}_{k})} \psi(\vect{r}_{k},t),
\label{eq:field_it4}
\end{eqnarray}
where $\module{\mu^2/(C_{1}C_{2}(\vect{r}))} = 1$ since $\module{\mu}
= \module{C_{1}} = \module{C_{2}(\vect{r})} = 1$.
 
It is now easy to convince ourselves that the form of \refp{eq:field_it4} is 
either $\partial^2\psi(\vec{r},t)/\partial t^2 = L\psi(\vec{r},t)$ 
or $i\partial\psi(\vec{r},t)/\partial t = L\psi(\vec{r},t)$, where $L$ 
is a real operator. Let us notice that the left hand side of 
equation \refp{eq:field_it4} contains the terms which are needed to build the 
time derivatives of the field. There are only two possibilities~: $\mu^2/
\left[C_{1}C_{2}(\vect{r})\right] = 1$ or $\mu^2/
\left[C_{1}C_{2}(\vect{r})\right] = -1$. If $\mu^2/
\left[C_{1}C_{2}(\vect{r})\right] = 1$, the left hand side of equation 
\refp{eq:field_it4} becomes $\psi(\vect{r},t+\tau)+\psi(\vect{r},t-\tau)$ 
which allows us to construct a second order time derivative~: $\left[
\psi(\vect{r},t+\tau)+\psi(\vect{r},t-\tau)-2\psi(\vect{r},t)\right]/\tau^2$. 
If $\mu^2/\left[C_{1}C_{2}(\vect{r})\right] = -1$, the difference 
$\psi(\vect{r},t+\tau)-\psi(\vect{r},t-\tau)$ is directly related to the 
first order time derivative~: $\left[\psi(\vect{r},t+\tau) -
\psi(\vect{r},t-\tau)\right]/2\tau$. The first or second order time 
derivatives of the field must be the same on the whole lattice, which is 
achieved by setting $C_{2}(\vect{r}) = C_{2}$. 
  
Lastly, we conclude that according to the two values we assign to 
$\mu^2/(C_{1}C_{2})$, equation \refp{eq:field_it4} becomes%

\[
\begin{array}{ccccc}
\displaystyle{\frac{\mu^2}{(C_{1}C_{2})}} = 1 & \rightarrow &
\psi(\vect{r},t+\tau) + \psi(\vect{r},t-\tau) & = & 
\displaystyle{\sum_{k=1}^{2d+1}}
\epsilon_{k}(\vect{r}) \module{\lambda_{k}(\vect{r})}
\module{\rho_{\overline{k}}(\vect{r}_{k})} \psi(\vect{r}_{k},t), \\
\\
\displaystyle{\frac{\mu^2}{(C_{1}C_{2})}} = -1 & \rightarrow & 
i\left[ \psi(\vect{r},t+\tau)-\psi(\vect{r},t-\tau)\right] & = &
\displaystyle{\sum_{k=1}^{2d+1}}
\epsilon_{k}(\vect{r}) \module{\lambda_{k}(\vect{r})}
\module{\rho_{\overline{k}}(\vect{r}_{k})} \psi(\vect{r}_{k},t),
\end{array} 
\]
as was claimed above. In particular, it is worthwhile to point out 
that, due to the square root of $\mu^2/(C_{1}C_{2})$ in the right hand side 
of \refp{eq:field_it4},  the factor $i$ appears when the equation is of 
first order in time.%
 
In the remainder of the paper, we shall write

\begin{equation}
\displaystyle{\frac{\mu^2}{(C_{1}C_{2})}} = \mu^2 e^{-i(\gamma_{1}+\gamma_{2})}
= \epsilon_{e},
\label{def:epsilon_e}
\end{equation}
with $\epsilon_{e}=\pm 1$.


\section{Reciprocity}~\label{re}
\begin{figure}[htb]
\begin{center}
\epsffile{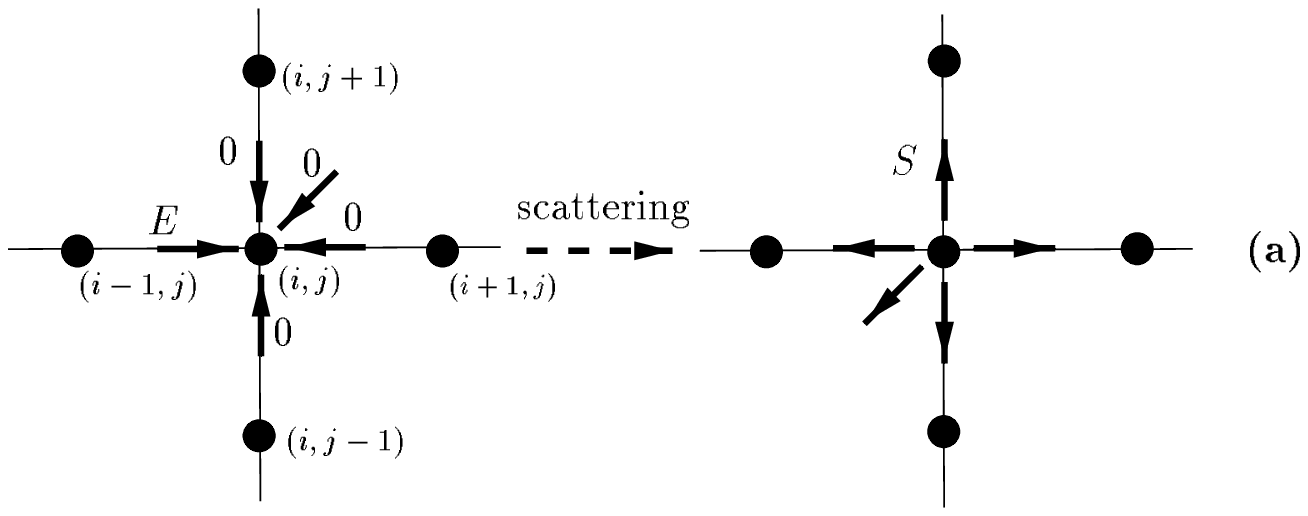}
\epsffile{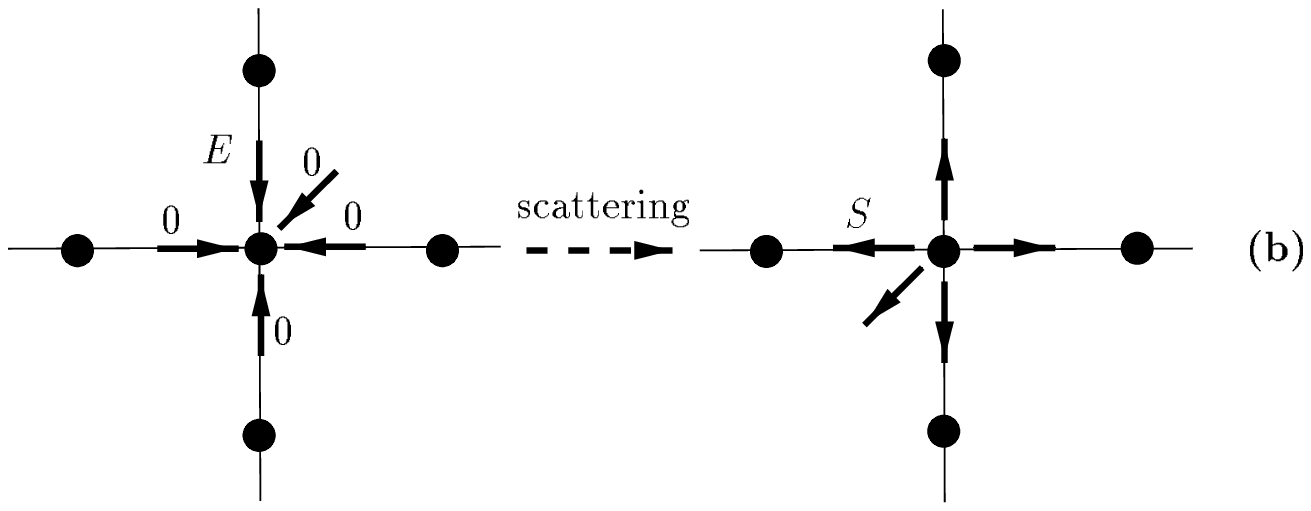}
\caption{Reciprocity. {\bf (a)} direct process. {\bf (b)} reciprocal 
process.}~\label{fig:re}
\end{center}
\end{figure}

In this section, we go further by imposing a new symmetry, reciprocity, 
on the evolution rules of the currents. The definition is sketched 
in Figure \ref{fig:re} where $E$ and $S$ are respectively 
an input and an outgoing current defined on two arbitrary bonds of 
the same node. We assume that the direct process in Fig.\ 
\ref{fig:re}~{\bf (a)} implies that the reciprocal 
process in Fig.\ \ref{fig:re}~{\bf (b)} also exists. 
In the reciprocal process, the currents $E$ and $S$ have the same 
values as in Figure \ref{fig:re}~{\bf (a)} but the two arbitrary bonds 
have been exchanged. 

Formulated in this 
way, this symmetry is the counterpart for the scattering process, 
of the reciprocity theorem which is well known for propagation 
in inhomogeneous media (see for 
instance\cite{landaua,landaub,economou,prada}). This is our 
main motivation for introducing it. Let us discuss its consequences. 

Using matrix notation, the process described in Fig.\ 
\ref{fig:re}~{\bf (a)} can be written 
as $S_{l}=s_{lk}E_{k}$ where $E_{k}$ is the input current and $S_{l}$ is any 
of the outgoing currents.  We can also imagine a second process for which 
the only non-vanishing input current is incident on bond $l$ while we are 
looking at the outgoing current on bond $k$, i.e. $S_{k}=s_{kl}E_{l}$. If 
both processes are chosen so that $E_{k}=E_{l}$, reciprocity tells us that 
$S_{k}=S_{l}$, leading to $s_{kl}=s_{lk}$. Since the two bonds $k$ and $l$ 
are arbitrary, one concludes that the scattering matrix is symmetrical

\[
S=S^{T}.
\] 
Additionaly, the scattering matrix is time-reversal invariant~: 
$S^{-1}=S^{\ast}$. Hence, the scattering matrix is unitary. As a direct 
consequence of this result, the sum of current intensities is conserved in 
any scattering process on the lattice

\begin{equation}
\sum_{k=1}^{2d+1} \module{S_{k}}^2 = \sum_{k=1}^{2d+1} \module{E_{k}}^2.
\label{eq:flux_cons}
\end{equation} 
Since the current intensities are also conserved in the propagation 
stage, their total sum over the whole lattice keeps the same 
value at each time step. This property guarantees the stability 
of the current dynamics. To understand its importance, let us 
consider a system of linear size $L$. The linear evolution of the 
whole set of currents can be described by a $(2d+1)L^d\times (2d+1)L^d$ 
matrix which transforms the $(2d+1)L^d$ input currents at time $t$ into the 
$(2d+1)L^d$ input currents at the next time step $t+\tau$. Such a linear 
system is characterized by its eigenvectors and the corresponding eigenvalues.
Any arbitrary initial state of the currents can be expanded on the basis of 
these eigenvectors. Let us choose one arbitrary eigenvector, characterized 
by its eigenvalue $\zeta$, as the initial state. The current intensities will 
diverge to infinity if $\module{\zeta}>1$ or decay to zero if 
$\module{\zeta}<1$. The above result, equation \refp{eq:flux_cons}, prevents 
us from such a scenario since the total sum of current intensities is 
constant regardless of the initial state. We conclude that all the 
eigenvalues are bound to be of modulus $1$ and that any initial linear 
superposition of eigenvectors is preserved by the time evolution of the 
currents. This result is pleasing because it is equivalent to what is expected 
from the acceptable solutions of the usual linear wave equations 
such as the classical wave or Schr\"{o}dinger equations. In order 
to break the stability of the current dynamics, losses or interactions 
of the system with the ``outside world'' must be introduced in 
some way.

Note that instead of introducing reciprocity, we could have required 
first the conservation law \refp{eq:flux_cons}, and then combining this law 
with time-reversal symmetry would have led to reciprocity.

From $s_{kl}=s_{lk}$ and \refp{eq:scatt4}, we find

\[
\rho_{k}\lambda_{l}=\rho_{l}\lambda_{k}, \esp \forall k,l,
\]
or

\[
\frac{\rho_{k}}{\lambda_{k}}=R e^{i\gamma},
\] 
where $R$ and $\gamma$ are independent of $k$.

Combining this result with the two first equations of \refp{eq:it_currents1} 
leads directly to $2\gamma=\gamma_{2}-\gamma_{1}$. From \refp{eq:it_currents2}
and the definition \refp{def:epsilon_e} of $\epsilon_e$, one obtains

\begin{equation}
\left\{
\begin{array}{ccc}
R & = & \displaystyle{\frac{\left(e^{i\gamma_{1}} + e^{-i\gamma_{2}}\right)
e^{i\gamma}}
{\sum_{k=1}^{2d+1} \module{\lambda_k}^2}}, \\
\\
\rho_{k} & = & \displaystyle{\frac{\left(1+\epsilon_{e}/\mu^2\right)
\mu_{k}\lambda_{k}^{\ast}}{\sum_{k=1}^{2d+1} \module{\lambda_{k}}^2}}.
\end{array}
\right.
\label{eq:rec1}
\end{equation} 
The matrix element becomes

\[
s_{kl} = \rho_{k}\lambda_{l}-\mu_{k}\delta_{kl} = \mu_{k}\left[ \displaystyle{
\frac{\left(1+\epsilon_{e}/\mu^2\right)\lambda_{k}^{\ast}\lambda_{l}}{
\sum_{k=1}^{2d+1} \module{\lambda_{k}}^2}} - \delta_{kl} \right].
\]
Remembering that, according to the second equation of \refp{eq:it_currents1},
$\mu_{k} = e^{i\gamma_{2}}\lambda_{k}/\lambda_{k}^{\ast}$, the scattering 
matrix is parametrized by two constant phases $\gamma_{1}$ and $\gamma_{2}$
and by $2d+1$ complex coefficients $\lambda_{k}$ which are functions of the 
location $\vect{r}$.


\section{Isotropy}~\label{is}
\begin{figure}[htb]
\begin{center}
\epsffile{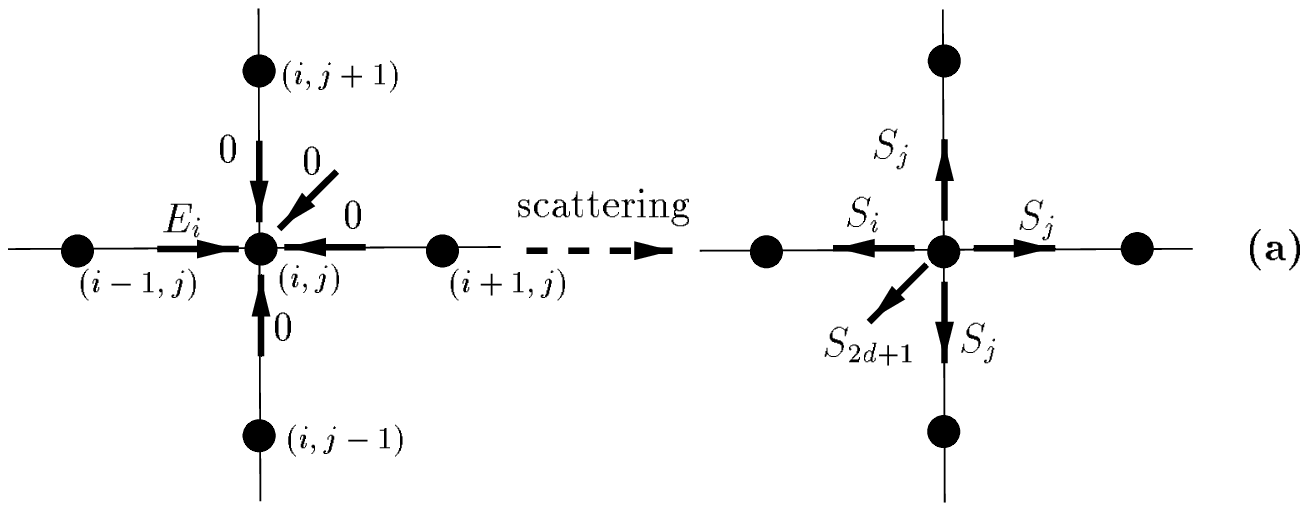}
\epsffile{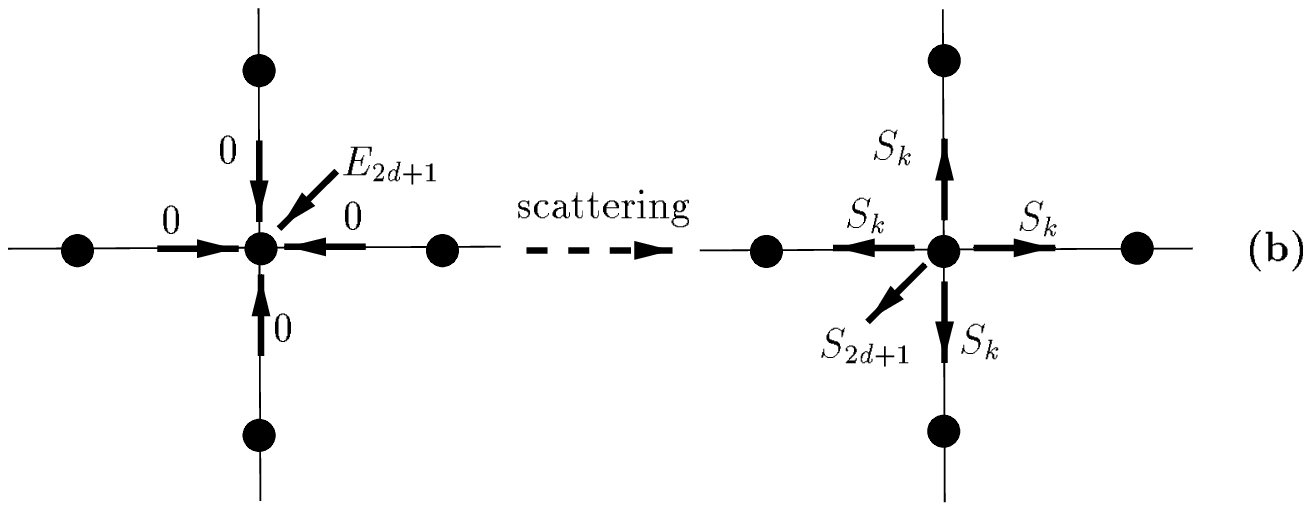}
\caption{The two scattering processes considered to 
define isotropy.}~\label{fig:isotropy}
\end{center}
\end{figure}

Isotropy is not required in order for the model to be solvable. 
However, we impose isotropy to simplify the model since anisotropic 
scatterers could be considered as well. Let us consider a process 
where only one of the propagating input currents $E_{i}$ is different from 
zero (Figure \ref{fig:isotropy}~{\bf (a)}). 
This process is by definition isotropic if all 
outgoing currents have the same value except for two outgoing currents~: 
$S_{i}$ on the input bond and the node current $S_{2d+1}$

\[
S_{j}=S_{k}, \esp \forall j,k \; /\; j,k \neq i, \; j,k \neq 2d+1. 
\]
In other words, when viewed from one particular bond, all other 
bonds are equivalent except the node bond because of its special 
status. The scatterer will be isotropic if the scattering process 
is isotropic for any chosen input bond.

Another process to consider is the one where the node current 
is different from zero (Figure \ref{fig:isotropy}~{\bf (b)}). 
This process is isotropic 
if all outgoing currents have the same value except the node 
current $S_{2d+1}$, i.e.

\[
S_{j}=S_{k}, \esp \forall j,k \; /\; j,k \neq 2d+1 
\]
Actually, both kinds of processes lead to the same conclusions because 
of the form of the scattering elements discussed in the previous 
section. 

\begin{equation}
\rho_{j}=\rho_{k}, \lambda_{j}=\lambda_{k}, \mu_{j}=\mu_{k}, \esp \forall
j,k \; /\; j,k \neq 2d+1.
\label{eq:iso}
\end{equation}

The scattering matrix now depends only on two complex parameters $\lambda_{1}$
and $\lambda_{2d+1}$, and on the two phases $\gamma_{1}$ and $\gamma_{2}$
(since $\mu_{k}$ and $\rho_{k}$ can be expressed as functions of $\lambda_{k}$ 
according to the second equations of \refp{eq:it_currents1} and 
\refp{eq:rec1}).

Conditions \refp{eq:iso} and \refp{eq:condsuf3} provide an additional 
constraint for the value of $\mu_{1}$. First, suppose the value of $\mu_{1}$ 
is known at node $\vect{r}$. Then, according to \refp{eq:condsuf3}, the value 
of $\mu_{\overline{k}}$ on the nearest-neighbors of node $\vect{r}$ reads
 
\[
\mu_{\overline{k}}(\vect{r}_{k}) = \displaystyle{\frac{\mu^2}
{\mu_{k}(\vect{r})}} = \displaystyle{\frac{\mu^2}{\mu_{1}(\vect{r})}}, \esp
\forall k \neq 2d+1.
\]
Since the scatterers $\vect{r}_{k}$ are also isotropic, 
$\mu_{\overline{k}}(\vect{r}_{k}) = \mu_{1}(\vect{r}_{k})$ for $k\neq 2d=1$, we
obtain

\[
\mu_{1}(\vect{r}_{k}) = \displaystyle{\frac{\mu^2}{\mu_{1}(\vect{r})}}, \esp
\forall k \neq 2d+1.
\]
Thus, $\mu_{1}(\vect{r}_{k})$ has a constant value on the nearest neighbors 
$\vect{r}_{k}$ of node $\vect{r}$. Next, starting from nodes $\vect{r}_{k}$, 
we obtain for the second nearest-neighbors $\vect{r}_{kl}$ of node $\vect{r}$,

\[
\mu_{1}(\vect{r}_{kl}) = \displaystyle{\frac{\mu^2}{\mu_{1}(\vect{r}_{k})}}=
\mu_{1}(\vect{r}).
\] 
Finally, proceeding in the same way from node to node, we find that the 
whole system is made of two intermingled Cartesian lattices characterized 
by their respective values $\mu_{1}$ and $\mu^2/\mu_{1}$. It is important to 
note that this result is peculiar to a Cartesian lattice, which has been 
considered from the beginning. Actually, the model we have discussed so far 
does not depend on the type of the discrete lattice. Indeed, if we had 
considered, for instance, a triangular lattice, a quasiperiodic lattice 
or even a random lattice (where the node connectivity changes from node 
to node, see Appendix), the time dependent equation 
of the field or the form of the scattering matrices would not have 
changed. However, the conclusion leading to the two values of 
$\mu_{1}$ on a Cartesian lattice would need to be modified on an 
arbitrary lattice.  Consider a triangular lattice in which the 
elementary cell contains three nodes connected by three bonds. 
Following an elementary closed loop of three bonds starting 
at node $\vect{r}$, we would find after a complete cycle

\[
\mu_{1}(\vect{r}) = \displaystyle{\frac{\mu^2}{\mu_{1}(\vect{r})}},
\] 
{\it i.e.}

\begin{equation}
\mu_{1}(\vect{r})=\epsilon_{1}\mu,
\label{eq:mu1}
\end{equation}
where $\epsilon_{1}=\pm 1$. Moreover, $\epsilon_{1}$ should have the same 
value at the three nodes and as a result over the whole system. As this 
conclusion is true for almost any arbitrary lattice, we are led to adopt 
\refp{eq:mu1} for the sake of generality. Remembering that $\mu$ is defined 
by its square $\mu^2$ in equation \refp{eq:condsuf3}, it is possible to 
choose once and for all $\epsilon_{1}=1$, i.e. $\mu_{1}=\mu$ 
 
Let us combine now these results with time reversal symmetry 
and reciprocity. For this purpose, we introduce the ratio 
$\nu=\lambda_{2d+1}/\lambda_{1}$. From the second equation of 
\refp{eq:it_currents1} and \refp{eq:condsuf4}, we write
 
\[
\mu_{1}=\mu_{2d+1}\displaystyle{\frac{\nu^{\ast}}{\nu}} 
= \epsilon_{2d+1}\mu \displaystyle{\frac{\nu^{\ast}}{\nu}}.
\]
Since $\mu_{1}=\mu$, we get

\[
\displaystyle{\frac{\nu^{\ast}}{\nu}} = \epsilon_{2d+1} = \pm 1.
\] 
That is, the ratio $\nu=\lambda_{2d+1}/\lambda_{1}$ is either real or 
imaginary.

On the other hand, reciprocity leads to $\rho_{2d+1}=\rho_{1}\nu$. Hence, 
$\rho_{2d+1}\lambda_{1}=\rho_{1}\lambda_{2d+1}=\rho_{1}\lambda_{1}\nu$ and 
$\rho_{2d+1}\lambda_{2d+1}=\rho_{1}\lambda_{1}\nu^2$. Therefore, 
the scattering matrix elements become

\begin{equation}
\left\{
\begin{array}{rcll}
s_{kl} & = & \rho_{1}\lambda_{1} - \mu\delta_{kl}, & \forall k,l \neq 2d+1, \\
\\
s_{kl} & = & s_{lk} = \rho_{1}\lambda_{1}\nu, & \forall k\neq 2d+1, l=2d+1, \\
\\
s_{kl} & = & \rho_{1}\lambda_{1}\nu^2 - \mu\epsilon_{2d+1}, & k=l=2d+1, 
\end{array}
\right.
\label{eq:scatt_iso}
\end{equation}  
where, using \refp{eq:rec1}

\begin{equation}
\rho_{1}\lambda_{1} = 
\displaystyle{\frac{\mu+\epsilon_{e}/\mu}{2d+\nu\nu^{\ast}}}. 
\label{eq:transmission}
\end{equation}
Let us make a few remarks about these last results. First, the 
coefficients $\lambda_{k}$ which appear in the expression of the field 
$\psi(\vect{r},t)=\sum_{k=1}^{2d+1} \lambda_{k}(\vect{r})E_{k}(\vect{r},t)$ 
are defined modulo a multiplicative constant. Obviously, the scattering 
matrix cannot depend on such an arbitrary constant, and indeed, one can see 
that it is only function of the ratio $\nu=\lambda_{2d+1}/\lambda_{1}$, 
according to \refp{eq:scatt_iso} and \refp{eq:transmission}. 
Next, $\mu$ and $\epsilon_{e}$ are constants defined over the whole system. 
Hence, $\nu$ is the sole quantity which depends upon the location $\vect{r}$ 
in the system (with the restriction $\nu^{\ast}/\nu=\epsilon_{2d+1}=\pm 1$). 
Given the definition $\nu=\lambda_{2d+1}/\lambda_{1}$, we see then that 
this is the node current which enables us to describe an inhomogeneous 
system as asserted before. A model without node currents, which corresponds 
to $\nu=0$, would have restrained us to an homogeneous system. We can 
interpret the node current as trapping a fraction of the wave energy before 
releasing it to neighboring nodes. The parameter $\nu(\vect{r})$ measures the 
local strength of this trapping effect. We shall see in the next section 
that the velocity of the wave is directly related to $\nu(\vect{r})$.
 
We can now  write the final form of the wave equation \refp{eq:field_it4} as

\[
\psi(\vect{r},t+\tau)+\epsilon_{e}\psi(\vect{r},t-\tau) = 
\lambda_{1}(\vect{r}) \left[ \sum_{k=1}^{2d} \rho_{1}(\vect{r}_{k})
\psi(\vect{r}_{k},t) + \nu^2(\vect{r})\rho_{1}(\vect{r})\psi(\vect{r},t)
\right].
\] 
Given this equation, we are led to introduce the new field

\[
\phi(\vect{r},t)=\rho_{1}(\vect{r})\psi(\vect{r},t)=\rho_{1}(\vect{r})
\lambda_{1}(\vect{r})\left[ \sum_{k=1}^{2d} E_{k}(\vect{r},t) + \nu(\vect{r})
E_{2d+1}(\vect{r},t) \right],
\] 
which obeys the following equation

\[
\phi(\vect{r},t+\tau)+\epsilon_{e}\phi(\vect{r},t-\tau)= \displaystyle{
\frac{\mu+\epsilon_{e}/\mu}{2d+\nu(\vect{r})\nu^{\ast}(\vect{r})}}
\left[ \sum_{k=1}^{2d}
\phi(\vect{r}_{k},t) + \nu^2(\vect{r})\phi(\vect{r},t) \right].
\] 
Like the scattering matrix elements \refp{eq:scatt_iso}, this equation depends 
only on the constants $\mu$ and $\epsilon_{e}$ and on the local parameter 
$\nu(\vect{r})$. We note that the coefficient $\lambda_{1}(\vect{r})$ comes 
into the expressions of the field, of the scattering matrix and of the wave 
equation as the product $\lambda_{1}(\vect{r})\rho_{1}(\vect{r})$. Therefore, 
it is possible and convenient to rename the quantities $\lambda_{1}(\vect{r})
\rho_{1}(\vect{r})$ and $\lambda_{2d+1}(\vect{r})\rho_{1}(\vect{r})$ as 
$\lambda_{1}(\vect{r})$ and $\lambda_{2d+1}(\vect{r})$ respectively. With 
this new notation, we can reformulate the above equations as

\begin{eqnarray}
\phi(\vect{r},t) & = & \lambda_{1}(\vect{r})\left[ \sum_{k=1}^{2d} 
E_{k}(\vect{r},t) + \nu(\vect{r})E_{2d+1}(\vect{r},t) \right], 
\label{eq:field} \\
\nonumber \\
\lambda_{1}(\vect{r}) & = & \displaystyle{\frac{\mu+\epsilon_{e}/\mu}
{2d+\nu(\vect{r})\nu^{\ast}(\vect{r})}},
\label{eq:coef_trans}
\end{eqnarray}

\begin{equation}
\phi(\vect{r},t+\tau) + \epsilon_{e}\phi(\vect{r},t-\tau) = 
\lambda_{1}(\vect{r}) \left[ \sum_{k=1}^{2d}\left( \phi(\vect{r}_{k},t)-
\phi(\vect{r},t) \right) + \left(2d+\nu^2(\vect{r})\right)\phi(\vect{r},t) 
\right],
\label{eq:we_1node}
\end{equation}
where we have shown explicitely the term 
$\sum_{k=1}^{2d}\left(\phi(\vect{r}_{k},t)-\phi(\vect{r},t)\right)$, which 
corresponds to the discrete Laplacian on an Euclidean lattice. Similarly, 
the scattering matrix is given by

\begin{equation}
S(\vect{r}) = \lambda_{1}(\vect{r})
\left(
\begin{array}{cccc}
1 & \cdots & 1 & \nu(\vect{r}) \\
\vdots & \vdots & \vdots & \vdots \\
\vdots & \vdots & \vdots & \vdots \\
1 & \cdots & 1 & \nu(\vect{r}) \\
\nu(\vect{r}) & \cdots & \nu(\vect{r}) & \nu^2(\vect{r}) 
\end{array}
\right)
-\mu
\left(
\begin{array}{ccccc}
1 & 0 & \cdots & 0 & 0 \\
0 & \ddots & \vdots & \vdots & 0 \\
\vdots & \vdots & \ddots & \vdots & \vdots \\
\vdots & \vdots & \vdots & 1 & \vdots \\
0 & \cdots & \cdots & 0 & \epsilon_{2d+1}(\vect{r}) \\
\end{array}
\right),
\label{eq:scatt_matrix}
\end{equation}
where $\epsilon_{2d+1}(\vect{r})=\nu^{\ast}(\vect{r})/\nu(\vect{r})=\pm 1$.
 

\section{Discussion of the wave equation}~\label{we}
The wave equation \refp{eq:we_1node} is a function of two discrete parameters, 
$\epsilon_{e}$ and $\epsilon_{2d+1}(\vect{r})$. The constant $\epsilon_{e}$ 
is defined over the whole system. We have already seen that its value $\pm 1$ 
controls the order of the time dependent wave equation, which will later 
be discussed separately. The parameter $\epsilon_{2d+1}(\vect{r})$ has a 
different satus. Being defined by 
$\epsilon_{2d+1}(\vect{r})=\nu^{\ast}(\vect{r})/\nu(\vect{r})$ with 
$\nu(\vect{r})$ controlling the disorder of the system, its value $\pm 1$ 
depends on the node $\vect{r}$. However, the type of disorder we would 
like to describe should depend continuously on the location in the system, 
such as a potential, a wave velocity or an index of refraction which varies 
smoothly with the position. Therefore, we shall assume that the value of 
$\epsilon_{2d+1}$ is fixed over the whole system whereas $\nu(\vect{r})$ 
is a smooth function of $\vect{r}$. If $\epsilon_{2d+1}=1$ or $-1$, 
$\nu(\vect{r})$ is real or imaginary, respectively. Both cases will also 
be considered separately. Hence, four possibilities will be discussed 
according to the values of $\epsilon_{e}$ and $\epsilon_{2d+1}$.

\subsection{Second order time equation $(\epsilon_{e}=1)$}~\label{so}
\subsubsection{$\epsilon_{2d+1}=1$}~\label{so1}
Since in this case $\nu(\vect{r})$ is real, 
$\nu(\vect{r})\nu^{\ast}(\vect{r})=\nu^2(\vect{r})$. The wave equation becomes

\[
\phi(\vect{r},t+\tau)+\phi(\vect{r},t-\tau)-2\phi(\vect{r},t)=
\displaystyle{\frac{2\cos\theta}{2d+\nu^2(\vect{r})}} 
\sum_{k=1}^{2d}\left(
\phi(\vect{r}_{k},t)-\phi(\vect{r},t)\right)+
2(\cos\theta - 1)\phi(\vect{r},t), 
\]
where we have written explicitly the second order time derivative and 
$\mu=e^{i\theta}$. 

We have already pointed out that our results were not 
peculiar to a Cartesian lattice. A similar wave equation would have been 
obtained in any discrete lattice. However, in order to consider 
this equation as the discretized version of the corresponding 
continuous wave equation, we need now to limit ourselves to a 
Cartesian lattice so that we can introduce the lattice constant $l$ and the 
current velocity $c_{0}=l/\tau$. Hence, one writes

\begin{eqnarray} 
\lefteqn{ \left(\displaystyle{\frac{2\cos\theta}
{2d+\nu^2(\vect{r})}}c_{0}^2\right)
\sum_{k=1}^{2d} \frac{1}{l^2}\left(\phi(\vect{r}_{k},t)-\phi(\vect{r},t)\right)}
\nonumber \\
&& -\frac{1}{\tau^2}\left(\phi(\vect{r},t+\tau)+\phi(\vect{r},t-\tau)-
2\phi(\vect{r},t)\right) = \frac{2(1-\cos\theta)}{\tau^2}\phi(\vect{r},t),
\label{eq:we_11}
\end{eqnarray}
which is the discretized version of the Klein-Gordon equation

\begin{equation}
c^2(\vect{r})\nabla^2\phi - \frac{\partial^2\phi}{\partial t^2} = a^2\phi.
\label{eq:kg}
\end{equation}
In the particular case where the Klein-Gordon equation describes a scalar 
particle of mass $m$, the coefficient $a$ would be given by 
$a^2=m^2c^4/\hbar^2$. Comparison between \refp{eq:we_11} and \refp{eq:kg} 
provides the velocity $c$ of the wave and the coefficient $a$

\begin{equation}
\left\{
\begin{array}{ccl}
c^2(\vect{r}) & = & c_{0}^2 \left(\displaystyle{\frac{2\cos\theta}
{2d+\module{\nu(\vect{r})}^2}}\right), \\
\\
a^2 & = & \displaystyle{\frac{2(1-\cos\theta)}{\tau^2}}.
\end{array}
\right.
\label{eq:coef_kg}
\end{equation} 
The special choice $\theta=0$, i.e. $\mu=1$, leads to the scalar wave 
equation $c^2(\vect{r})\nabla^2\phi - \partial^2\phi/\partial t^2 = 0$. 
The first equation of \refp{eq:coef_kg} shows clearly that the velocity of 
the wave is directly connected to the parameter $\nu(\vect{r})$. Its maximum 
value $c_{max}=c_{0}(\cos\theta/d)^{1/2}$ is reached when 
$\nu(\vect{r})=0$. As stated before, the role of the node current is to 
slow down the wave. If $d>1$, we also note that $c_{max}$ is always 
strictly smaller than the velocity $c_{0}$ of the currents.

\subsubsection{$\epsilon_{2d+1}=-1$}
In this case $\nu(\vect{r})$ is imaginary, thus 
$\nu^2(\vect{r})=-\module{\nu(\vect{r})}^2$. The wave equation becomes
 
\begin{eqnarray}
\lefteqn{ \left(\displaystyle{\frac{2\cos\theta}
{2d+\module{\nu(\vect{r})}^2}}c_{0}^2\right)
\sum_{k=1}^{2d} \frac{1}{l^2}\left(\phi(\vect{r}_{k},t)-\phi(\vect{r},t)\right)}
\nonumber \\
&& -\frac{1}{\tau^2}\left(\phi(\vect{r},t+\tau)+\phi(\vect{r},t-\tau)-
2\phi(\vect{r},t)\right) =\frac{2}{\tau^2}\left[
\cos\theta \left(\displaystyle{\frac{2d-\module{\nu(\vect{r})}^2}
{2d+\module{\nu(\vect{r})}^2}}\right)+1 \right]\phi(\vect{r},t),
\label{eq:we_1-1}
\end{eqnarray}
This equation is also a discretized version of \refp{eq:kg} since the factor 
of $\phi(\vect{r},t)$ in the right hand side of \refp{eq:we_1-1} is positive. 
The velocity of the wave is given by

\[
c^2(\vect{r}) = c_{0}^2 \left(\displaystyle{\frac{2\cos\theta}
{2d+\module{\nu(\vect{r})}^2}}\right).
\]
However, \refp{eq:we_1-1} is not as versatile as \refp{eq:we_11}. First, 
in order to obtain the scalar wave equation, we must set the double condition 
$\theta=0$ and $\nu(\vect{r})=0$. Hence, the wave velocity is pinned to its 
maximum value $c=c_{0}/\sqrt{d}$. This describes a uniform system without 
node current. Next, it is easy to show that the same double condition holds 
if we take the continuous limit \refp{eq:we_1-1} in fixing the value of $a^2$. 
Under these circumstances, we conclude that an imaginary node current is not 
compatible with equation \refp{eq:kg}.

Hence, from the analysis of these two cases $(\epsilon_{2d+1}=\pm 1)$, 
in order to correctly describe the Klein-Gordon equation and 
the scalar wave equation in a inhomogeneous system by our current 
model, we must set $\epsilon_{e}=1$ and $\epsilon_{2d+1}=1$. The resulting 
equation is given by \refp{eq:we_11}. This result is not completely 
satisfactory since the necessity of setting $\epsilon_{2d+1}=1$ 
is not derived from the general kind of reasoning which has been 
used  in our approach up to this stage. We will see in section 
\ref{tn} that adding a second node current to each scatterer will remove 
this arbitrary condition.

\subsection{First order time equation $(\epsilon_{e}=-1)$}~\label{fo}
\subsubsection{$\epsilon_{2d+1}=1$} 
With $\nu(\vect{r})$ being real, the wave equation becomes

\[
\frac{i}{2\tau}\left(\phi(\vect{r},t+\tau)-\phi(\vect{r},t-\tau)\right)=
-\frac{l^2}{\tau}\left(\frac{\sin\theta}{2d+\nu^2(\vect{r})}\right) 
\sum_{k=1}^{2d}\left(
\displaystyle{\frac{\phi(\vect{r}_{k},t)-\phi(\vect{r},t)}{l^2}} \right)
-\frac{\sin\theta}{\tau}\phi(\vect{r},t).
\]
We recognize the discretized version of the Schr\"{o}dinger equation

\begin{equation}
i\hbar\frac{\partial\phi}{\partial t} = -\frac{\hbar^2}{2m}\nabla^2\phi + 
V(\vect{r})\phi,
\label{eq:schrodinger}
\end{equation}
with

\[
\frac{V(\vect{r})}{\hbar} = -\frac{\sin\theta}{\tau},
\] 
and

\[
\frac{\hbar}{2m} = \frac{l^2}{\tau} \left( 
\frac{\sin\theta}{2d+\module{\nu(\vect{r})}^2} \right) =
-\frac{l^2 V(\vect{r})}{\hbar(2d+\module{\nu(\vect{r})}^2)}.
\] 
The expression for $V(\vect{r})$ shows that we would obtain a 
Schr\"{o}dinger equation in which the potential is bound to stay constant 
over the whole system. Therefore, we shall reject the choice 
$\epsilon_{e}=-\epsilon_{2d+1}=-1$.

\subsubsection{$\epsilon_{2d+1}=-1$}
Again, $\nu^2(\vect{r})=-\module{\nu(\vect{r})}^2$. We obtain

\begin{eqnarray}
\lefteqn{ \frac{i}{2\tau}\left(\phi(\vect{r},t+\tau)-
\phi(\vect{r},t-\tau)\right)=} \nonumber \\
&& -\frac{l^2}{\tau} \left( 
\frac{\sin\theta}{2d+\module{\nu(\vect{r})}^2} \right) 
\sum_{k=1}^{2d}\left(
\displaystyle{\frac{\phi(\vect{r}_{k},t)-\phi(\vect{r},t)}{l^2}} \right)
-\frac{\sin\theta}{\tau}\left( \displaystyle{
\frac{2d-\module{\nu(\vect{r})}^2}{2d+\module{\nu(\vect{r})}^2}} \right)
\phi(\vect{r},t). 
\label{eq:we_-1-1}
\end{eqnarray}
In a similar way, identification with the discrete Schr\"{o}dinger 
equation \refp{eq:schrodinger} gives

\[
\frac{V(\vect{r})}{\hbar} = -\frac{\sin\theta}{\tau} \left(\displaystyle{
\frac{2d-\module{\nu(\vect{r})}^2}{2d+\module{\nu(\vect{r})}^2}}\right),
\]
and 

\[
\frac{\hbar}{2m} = \frac{l^2}{\tau} \left(\frac{\sin\theta}{2d+
\module{\nu(\vect{r})}^2}\right) =
-\frac{l^2 V(\vect{r})}{\hbar(2d-\module{\nu(\vect{r})}^2)}.
\] 
In contrast with the previous section, the potentiel $V(\vect{r})$ is a 
function of the position $\vect{r}$. The last equality shows that the mass $m$ 
also depends on $\vect{r}$. This is not a surprise in the context of quantum 
waves on a lattice where one usually introduces an effective mass which 
may be a function of position (see\cite{sheng} for instance). However, the 
Schr\"{o}dinger equation \refp{eq:we_-1-1} is not completely general. 
We note that the variations of $m$ and $V$ with $\vect{r}$ are correlated. 
In particular, the mass cannot be kept fixed at a constant value if the 
potential is a function of $\vect{r}$. In the context of the localization 
of quantum waves in an inhomogeneous medium, equation \refp{eq:we_-1-1} 
would correspond to the Anderson model with correlated ``diagonal'' and 
``off-diagonal'' disorder\cite{sheng}. Hence, one may wonder 
which conditions must be imposed in our approach in order 
to derive a Schr\"{o}dinger equation where the two types of 
disorder can be described separately. 
The most natural idea which comes to mind, is that one needs 
an additionnal degree of freedom which would allow to separate 
the mass and the potential spatial variations. Introducing a second 
node current not only achieves this task but also removes the 
arbitrary condition discussed in section~\ref{so}.


\section{Introducing a second node current}~\label{tn}
Let us attach to each node additional currents $E_{2d+2}$ and $S_{2d+2}$ 
which behave like the node currents $E_{2d+1}$ and $S_{2d+1}$. They 
participate in the same scattering process as the other currents but do not 
propagate to neighbor nodes. Instead, the outgoing current $S_{2d+2}$ becomes 
the incident current $E_{2d+2}$ on the same node at the next time step. 
The scattering process and the definition of the field become

\begin{eqnarray*}
S_{k} & = & \sum_{l=1}^{2d+2}s_{kl}E_{l}, \esp k=1,\ldots,2d+2, \\
\psi(\vect{r},t) & = & \sum_{k=1}^{2d+2} \lambda_{k}(\vect{r}) E_{k}(\vect{r}).
\end{eqnarray*} 
It is easy to check that the derivation of a wave equation similar 
to \refp{eq:we_1node} remains quite unchanged. Therefore, we will only point 
out the most noticeable differences and the final result.

First, equation \refp{eq:condsuf4} is supplemented by

\[
\mu_{2d+2}(\vect{r}) = \epsilon_{2d+2}(\vect{r})\mu,
\]
where $\epsilon_{2d+2}(\vect{r})=\pm 1$.

Second, instead of the single ratio $\nu=\lambda_{2d+1}/\lambda_{1}$, 
we introduce $\nu_{2d+1}=\lambda_{2d+1}/\lambda_{1}$ and $\nu_{2d+2}=
\lambda_{2d+2}/\lambda_{1}$ which obey the conditions 
$\nu^{\ast}_{2d+1}/\nu_{2d+1}=\epsilon_{2d+1}=\pm 1$ and 
$\nu^{\ast}_{2d+2}/\nu_{2d+2}=\epsilon_{2d+2}=\pm 1$. In other words, we find 
again that $\nu_{2d+1}$ and $\nu_{2d+2}$ are either real or imaginary.

Finally, equations \refp{eq:field}, \refp{eq:coef_trans} and 
\refp{eq:we_1node} become

\begin{eqnarray}
\phi(\vect{r},t) & = & \lambda_{1}(\vect{r})\left[ \sum_{k=1}^{2d} 
E_{k}(\vect{r},t) + \nu_{2d+1}(\vect{r})E_{2d+1}(\vect{r},t)
+ \nu_{2d+2}(\vect{r})E_{2d+2}(\vect{r},t) \right], 
\label{eq:field_2node} \\
\nonumber \\
\lambda_{1}(\vect{r}) & = & \displaystyle{\frac{\mu+\epsilon_{e}/\mu}
{2d+\module{\nu_{2d+1}(\vect{r})}^2+\module{\nu_{2d+2}(\vect{r})}^2}},
\label{eq:coef_trans_2node}
\end{eqnarray}

\begin{eqnarray}
\lefteqn{\phi(\vect{r},t+\tau) + \epsilon_{e}\phi(\vect{r},t-\tau) =}
\nonumber \\ 
& &	\lambda_{1}(\vect{r}) \left[ 
	\sum_{k=1}^{2d}\left( \phi(\vect{r}_{k},t)-
	\phi(\vect{r},t) \right) + \left(2d+\nu_{2d+1}^2(\vect{r})
	+\nu_{2d+2}^2(\vect{r})\right)\phi(\vect{r},t) \right].
\label{eq:we_2node}
\end{eqnarray}
At this stage, the discussion of section~\ref{we} can be reproduced. 
However, as
shown below, we need only to discuss the two cases $\epsilon_{e}=\pm 1$
separately. Indeed, we have been previously led to discuss the 
two choices $\nu=\lambda_{2d+1}/\lambda_{1}$ being real or imaginary. 
By adding a second node current, it happens now that we have gained 
additional possibilities since $\nu_{2d+1}=\lambda_{2d+1}/\lambda_{1}$
and $\nu_{2d+2}=\lambda_{2d+2}/\lambda_{1}$ can be real or 
imaginary independently of each other. However, one can easily 
convince oneself that choosing $\nu_{2d+1}$ and $\nu_{2d+2}$ being 
simultaneously real or imaginary is equivalent to having a single 
node current such that $\nu^2=\nu^2_{2d+1}+\nu^2_{2d+2}$. The only new 
possibility opened up by introducing two nodes 
currents corresponds to $\nu_{2d+1}$ real and $\nu_{2d+2}$ imaginary, or 
{\it vice-versa}. Since both node currents have been equivalent until now, 
it is sufficient to consider one of both cases, i.e $\nu_{2d+1}$ real 
$(\epsilon_{2d+1}(\vect{r})=1)$ and $\nu_{2d+2}$ imaginary 
$(\epsilon_{2d+2}(\vect{r})=-1)$.

\subsection{Second order time equation $(\epsilon_{e}=1)$}
By introducing the lattice constant $l$ and the current velocity 
$c_{0}=l/\tau$, the wave equation \refp{eq:we_2node} becomes

\begin{eqnarray}
\lefteqn{ \left(\displaystyle{\frac{2\cos\theta}{2d+
\module{\nu_{2d+1}(\vect{r})}^2+\module{\nu_{2d+2}(\vect{r})}^2}}c_{0}^2\right)
\sum_{k=1}^{2d} \frac{1}{l^2}\left(\phi(\vect{r}_{k},t)-\phi(\vect{r},t)
\right)} \nonumber \\
& &	-\frac{1}{\tau^2}\left(\phi(\vect{r},t+\tau)+\phi(\vect{r},t-\tau)-
	2\phi(\vect{r},t)\right) = \nonumber \\
& & 	\frac{2}{\tau^2}\left[
	\cos\theta \left(\displaystyle{
\frac{2d+\module{\nu_{2d+1}(\vect{r})}^2-\module{\nu_{2d+2}(\vect{r})}^2}
	{2d+\module{\nu_{2d+1}(\vect{r})}^2
	+\module{\nu_{2d+2}(\vect{r})}^2}}\right)+1\right]\phi(\vect{r},t),
\label{eq:we_kg}
\end{eqnarray} 
where we have written $\mu=e^{i\theta}$.

Comparison between \refp{eq:we_kg} and the Klein-Gordon equation \refp{eq:kg} 
provides

\[
\left\{
\begin{array}{ccl}
c^2(\vect{r}) & = & c_{0}^2\left(\displaystyle{\frac{2\cos\theta}
{2d+\module{\nu_{2d+1}(\vect{r})}^2+\module{\nu_{2d+2}(\vect{r})}^2}}\right),\\
\\
a^2(\vect{r}) & = & \displaystyle{\frac{2}{\tau^2}}
\left[
\cos\theta \left(\displaystyle{\frac{2d+\module{\nu_{2d+1}(\vect{r})}^2-
\module{\nu_{2d+2}(\vect{r})}^2}
{2d+\module{\nu_{2d+1}(\vect{r})}^2+\module{\nu_{2d+2}(\vect{r})}^2}}\right)+1 
\right].
\end{array}
\right.
\]
These results are an improvement over those obtained in section~\ref{so1} 
since $a^2$ is a function of the position $\vect{r}$. Moreover, 
$c^2(\vect{r})$ and $a^2(\vect{r})$ can vary independently from each other 
by properly adjusting the variations of $\nu_{2d+1}(\vect{r})$ and 
$\nu_{2d+2}(\vect{r})$. Finally, the necessity of setting $\epsilon_{2d+1}=1$
which was found quite arbitrary in section~\ref{so}, is no longer 
pertinent. The above results show that one really needs to keep 
both $\epsilon_{2d+1}=1$ and $\epsilon_{2d+2}=-1$ simultaneously in order 
to obtain a complete version of the Klein-Gordon equation.

\subsection{First order time equation $(\epsilon_{e}=-1)$}
The wave equation \refp{eq:we_2node} becomes

\begin{eqnarray}
\lefteqn{ \frac{i}{2\tau}\left(\phi(\vect{r},t+\tau)-
\phi(\vect{r},t-\tau)\right)=} \nonumber \\
&& -\frac{l^2}{\tau} \left( 
\frac{\sin\theta}{2d+\module{\nu_{2d+1}(\vect{r})}^2
+\module{\nu_{2d+2}(\vect{r})}^2} \right) \sum_{k=1}^{2d}\left(
\displaystyle{\frac{\phi(\vect{r}_{k},t)-\phi(\vect{r},t)}{l^2}} \right)
\nonumber \\
&& -\frac{\sin\theta}{\tau}\left( \displaystyle{
\frac{2d+\module{\nu_{2d+1}(\vect{r})}^2-\module{\nu_{2d+2}(\vect{r})}^2}
{2d+\module{\nu_{2d+1}(\vect{r})}^2+\module{\nu_{2d+2}(\vect{r})}^2}} \right)
\phi(\vect{r},t). 
\label{eq:we_schrod}
\end{eqnarray}
Comparison between \refp{eq:we_schrod} and the Schr\"{o}dinger equation 
\refp{eq:schrodinger} gives
 
\[
\frac{V(\vect{r})}{\hbar} = -\frac{\sin\theta}{\tau} \left(\displaystyle{
\frac{2d+\module{\nu_{2d+1}(\vect{r})}^2-\module{\nu_{2d+2}(\vect{r})}^2}
{2d+\module{\nu_{2d+1}(\vect{r})}^2+\module{\nu_{2d+2}(\vect{r})}^2}}\right),
\]
and

\[
\frac{\hbar}{2m} = \frac{l^2}{\tau} \left(\displaystyle{ \frac{\sin\theta}{2d+
\module{\nu_{2d+1}(\vect{r})}^2+\module{\nu_{2d+2}(\vect{r})}^2}}\right). 
\]  
In contrast with the results discussed in section~\ref{fo}, this new 
version of the discrete Schr\"{o}dinger equation is now well 
defined. The potential $V(\vect{r})$ and the mass $m(\vect{r})$ can vary 
independently from each other by properly adjusting the variations of 
$\nu_{2d+1}(\vect{r})$ and $\nu_{2d+2}(\vect{r})$. For instance, it is 
possible to formulate the Anderson model with diagonal and 
off-diagonal disorders separately. Of particular interest is the choice 
$\module{\nu_{2d+1}(\vect{r})}^2+\module{\nu_{2d+2}(\vect{r})}^2 =
constant$ which corresponds to keeping $m(\vect{r}) = constant$
while $V(\vect{r})$ is varying with $\nu_{2d+1}(\vect{r})$ and 
$\nu_{2d+2}(\vect{r})$. 
  
Introducing two node currents could seem at first glance somehow 
arbitrary but is justified by the resulting well-defined Klein-Gordon 
and  Schr\"{o}dinger equations, equations \refp{eq:we_kg} and 
\refp{eq:we_schrod} respectively. 
These equations do not suffer from the drawbacks of the $2d+1$ 
current model discussed in section~\ref{we}.


\section{Conclusion}~\label{co}
It is at first glance a curiosity that starting from a discrete 
formulation of Huygens' principle and taking account of basic 
properties and symmetries, we have derived two equations that 
can be directly identified with the Klein-Gordon equation and 
the Schr\"{o}dinger equation. On the one hand, this is not completely 
surprising considering that this simple model incorporates the 
basic features of wave propagation. On the other hand, the temporal 
and spatial symmetries which have been selected in building the 
model are simply those underlying the resulting discrete wave 
equations. The novelty of this approach is the systematic derivation 
resulting in a versatile unified equation, which properly tuned 
by a single parameter, yields two of the most fundamental wave 
equations in physics. So this derivation can be viewed as a general 
framework into which is merged the related approach of the TLM 
network model. In particular, it is worthwhile to point out that 
this unified equation can be derived on any discrete lattice 
or graph, thus underlining the generality of the construction 
(see Appendix). 
  
The most unusual feature of this formulation is the introduction 
of currents without any initial physical meaning. This can be 
compared with the cellular automata approach which is now routinely 
used for solving hydrodynamic problems. By starting from a minimal 
microscopic model of non physical particles which obey the most 
fundamental conservation laws of physics such as energy and momentum 
conservation, the macroscopic behavior of real fluids is recovered. 
The current model described in this paper has a similar status. 
A physical interpretation of the currents was given in the TLM 
network model by identifying them with electric field impulses 
traveling on a mesh of transmission lines\cite{johns71}. Since in that 
particular case it can be shown that the voltages and currents 
obey the same equations as the electric and magnetic fields of 
Maxwell's equations, it is not surprising to obtain Maxwell's 
equations from that model. However, such an interpretation fails 
in the abstract context we have considered in this paper, especially 
when we obtain the Schr\"{o}dinger equation without any explicit 
analytic continuation from the diffusion equation (e.g. 
$t \mapsto it$\cite{enders95}). 
Although they locally propagate as particles, and then, are scattered 
as waves, the currents cannot be interpreted either as fields 
or as particles. This is also in contrast with other approaches 
where the Schr\"{o}dinger and Maxwell equations have been obtained 
from the dynamics of ensembles of Brownian 
particles\cite{ord96b,ord96a,ord97}, 
thus providing a particle picture to standard wave equations.
 
In considering further investigations, the natural question that 
comes immediatly to mind, concerns the possibility of describing 
other wave equations by the same approach. Some clues have already 
been indicated in this paper. For instance, we have discussed 
the possibility of describing higher order time derivatives of 
the field by releasing condition \refp{eq:condsuf3}. More generally, as the 
construction relies essentially on symmetries, we may wonder 
what equations would be obtained if a symmetry condition is not 
fulfilled or is replaced by another one. For instance, it will 
be easy to release the time-reversal symmetry and to solve numerically 
the Schr\"{o}dinger equation within this assumption. This would 
be of fundamental interest in our understanding of strong localization 
in random media. Finally, there is no reason for limiting ourselves 
to scalar equations. As discussed in the introduction, the Maxwell 
equations as an example of vector wave equations, have been exhibited 
in the TLM approach. A further possibility is the description 
of spinor waves by introducing two different types of currents 
propagating on a Cartesian lattice. This work will be described 
in a forthcoming paper\cite{sdetoro}.
  

\acknowledgements

The authors wish to thank O.{\sc Legrand}, F.{\sc Mortessagne}, 
P.{\sc Sebbah} and D.{\sc Sornette} for useful discussions on topics 
related to this work. They are also indebted to J.M.{\sc Luck} and 
M.{\sc Sundheimer} for a critical reading of the manuscript. 


\appendix
\section*{Formulation of the current-model on a graph}~\label{ap}
Although a $n$-dimensional Cartesian lattice has been considered 
throughout this paper,  we have pointed out  that most of our 
results did not depend on this particular lattice. This special 
kind of lattice, which involves an underlying manifold equipped 
with a metric, is required when considering the continuous limit 
of the discrete wave equation \refp{eq:we_2node}. Indeed, in that limit, the 
lattice spacing $l$ and the current velocity $c_{0}=l/\tau$ are needed. 
In this appendix, we show that our model leads to the discrete wave 
equation \refp{eq:we_2node} formulated on any arbitrary lattice. 
We are not going to reproduce in detail the derivation of 
Eq.\ \refp{eq:we_2node}, which remains quite unchanged. Instead of this, 
we focus on the main differences between both approaches on either a Cartesian 
lattice or on any arbitrary lattice.

\begin{figure}[htb]
\begin{center}
\epsffile{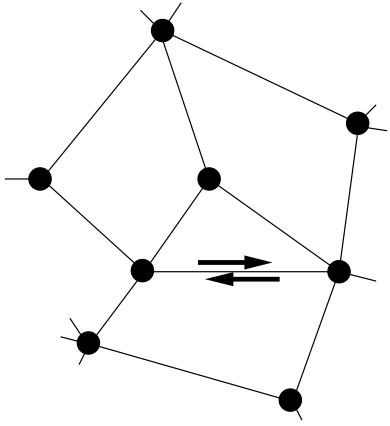}
\caption{An arbitrary discrete lattice.}~\label{fig:affine}
\end{center}
\end{figure}

We consider the most general lattice defined as a random lattice 
where each node is connected to $z(\vect{r})$ neighbor nodes, $z(\vect{r})$ 
denotes the coordination number. The notation $z(\vect{r})$ indicates 
explicitly that the number of neighbors varies from node to node, 
as sketched in Fig.\ \ref{fig:affine}. It is important to emphasize 
that no length is attributed to the bonds linking any two nodes. 
Thus, the bonds connected to one node represent only the connectivity 
of this arbitrary node. 

As explained in the introduction, the 
currents propagate in one time step between a pair of nodes of a given bond. 
By taking into account the node currents, $z(\vect{r})+2$ currents 
$E_{k}(\vect{r},t)$ are incident on node $\vect{r}$ at each time step. 
Hence, a $(z(\vect{r})+2)\times (z(\vect{r})+2)$ scattering matrix is 
attached to node $\vect{r}$. The definition \refp{def:field} of the field \
becomes

\begin{equation}
\psi(\vect{r},t) = \sum_{k=1}^{z(\smvec{r})+2} \lambda_{k}(\vect{r})
E_{k}(\vect{r},t).
\label{def:field_graph}
\end{equation} 
The number of currents in the definition of the field depends now 
on the node $\vect{r}$. Keeping in mind these definitions, and using 
the previous notation for the neighbors of node $\vect{r}$, as described 
in Figure \ref{fig:def_notbond}, it is easy 
to check that all the steps of the derivation 
of equation \refp{eq:we_2node} are left unchanged. 
For instance, conditions \refp{eq:scatt3} and \refp{eq:condsuf3} still hold 
to derive a closed field equation analogous to equation \refp{eq:field_out5}. 
The new  closed field equation reads straightforwardly

\begin{equation}
\psi(\vect{r},t+\tau) = \sum_{k=1}^{z(\smvec{r})+2} \lambda_{k}(\vect{r})
\rho_{\overline{k}}(\vect{r}_{k})\psi(\vect{r}_{k},t) + 
\mu^2 \left[ 1 - \sum_{k=1}^{z(\smvec{r})+2} 
\displaystyle{\frac{\lambda_{k}(\vect{r})\rho_{k}(\vect{r})}
{\mu_{k}(\vect{r})}} \right] \psi(\vect{r},t-\tau). 
\label{eq:we_graph}
\end{equation}
Then, the implementation of time-reversal symmetry and reciprocity 
can exactly be reproduced by substituting each summation over $2d+2$ currents 
by a summation over $z(\vect{r})+2$ currents. Isotropy is also introduced in 
the same way by noticing again that the number of propagating currents 
involved in this local symmetry now depends on the node $\vect{r}$. 
As discussed in section \ref{is}, the lattice being arbitrary, 
equation \refp{eq:mu1} holds and the sign $\epsilon_{1}$ can be fixed equal 
to $1$. Then, we introduce naturally the ratios 
$\nu_{z(\smvec{r})+1}(\vect{r})=\lambda_{z(\smvec{r})+1}/\lambda_{1}$ and 
$\nu_{z(\smvec{r})+2}(\vect{r})=\lambda_{z(\smvec{r})+2}/\lambda_{1}$ 
which satisfy the conditions 
$\nu^{\ast}_{z(\smvec{r})+1}(\vect{r})/\nu_{z(\smvec{r})+1}(\vect{r})=
\epsilon_{z(\smvec{r})+1}=\pm 1$ and  
$\nu^{\ast}_{z(\smvec{r})+2}(\vect{r})/\nu_{z(\smvec{r})+2}(\vect{r})=
\epsilon_{z(\smvec{r})+2}=\pm 1$. Finally, the equations 
\refp{eq:field_2node}, \refp{eq:coef_trans_2node} and 
\refp{eq:we_2node} are recovered in this context and read%
  
\begin{eqnarray}
\phi(\vect{r},t) & = & \lambda_{1}(\vect{r})\left[ \sum_{k=1}^{z(\smvec{r})} 
E_{k}(\vect{r},t) + \nu_{z(\smvec{r})+1}(\vect{r})
E_{z(\smvec{r})+1}(\vect{r},t)+\nu_{z(\smvec{r})+2}(\vect{r})
E_{z(\smvec{r})+2}(\vect{r},t) \right], 
\label{eq:field_2node_graph} \\
\nonumber \\
\lambda_{1}(\vect{r}) & = & \displaystyle{\frac{\mu+\epsilon_{e}/\mu}
{z(\vect{r})+\module{\nu_{z(\smvec{r})+1}(\vect{r})}^2
+\module{\nu_{z(\smvec{r})+2}(\vect{r})}^2}},
\label{eq:coef_trans_2node_graph}
\end{eqnarray}

\begin{eqnarray}
\lefteqn{ \phi(\vect{r},t+\tau) + \epsilon_{e}\phi(\vect{r},t-\tau) =} 
\nonumber \\ 
& &	\lambda_{1}(\vect{r}) \left[ \sum_{k=1}^{z(\smvec{r})}\left( 
	\phi(\vect{r}_{k},t)-\phi(\vect{r},t) \right) + 
	\left(z(\vect{r})+\nu_{z(\smvec{r})+1}^2(\vect{r})
	+\nu_{z(\smvec{r})+2}^2(\vect{r})\right)\phi(\vect{r},t) \right],
\label{eq:we_2node_graph}
\end{eqnarray}
where $\epsilon_{e}=\pm 1$ and $\mu=e^{i\theta}$.
 
In equation \refp{eq:we_2node_graph}, one recognizes the term 
$\sum_{k=1}^{z(\smvec{r})}\left( 
\phi(\vect{r}_{k},t)-\phi(\vect{r},t) \right)$ as a topological Laplacian 
defined on a graph. Thus, equation \refp{eq:we_2node_graph} is a wave 
propagation equation formulated for a scalar field defined on a graph. 
This result can be compared to the attempts of formulating 
discretized field theories for scalar fields\cite{regge,itzykson}.
 

\end{document}